\documentclass[a4paper]{article}

\usepackage[dvipdfmx]{graphicx, color}
\usepackage{wrapfig}
\usepackage{amsmath, amssymb}
\usepackage{bm}
\usepackage{pifont}
\usepackage{cite}
\usepackage[dviout]{pict2e}
\usepackage{booktabs}

\begin{document}

\title{Evaluation of measurement accuracy of $h \to \tau ^+ \tau ^-$ branching ratio \\
at the ILC with $\sqrt{s} = 250$ GeV and 500 GeV}
\date{\today}
\author{Shin-ichi Kawada$^{1,\dagger}$, Keisuke Fujii$^2$, Taikan Suehara$^3$,\\ Tohru Takahashi$^1$, Tomohiko Tanabe$^4$ }
\maketitle{}

\noindent 1: Advanced Sciences of Matter (AdSM), Hiroshima University, 1-3-1, Kagamiyama, Higashi-Hiroshima, Hiroshima, 739-8530, Japan \\
2: High Energy Accelerator Research Organization (KEK), 1-1, Oho, Tsukuba, Ibaraki, 305-0801, Japan \\
3: Department of Physics, Tohoku University, 6-3, Aoba, Aramaki, Aoba-ku, Sendai, Miyagi, 980-8578, Japan \\
4: International Center for Elementary Particle Physics (ICEPP), The University of Tokyo, 7-3-1, Hongo, Bunkyo-ku, Tokyo, 113-0033, Japan \\ \\
$\dagger$ : \verb|s-kawada@huhep.org|

\begin{abstract}
\footnote{This write-up is intended to supplement a white paper on the Higgs physics at the ILC.
A white paper on the Higgs physics at the ILC to be submitted to the Snowmass process 2013.}
We evaluate the measurement accuracy of the branching ratio of $h \to \tau ^+ \tau ^-$ at $\sqrt{s} = 250$ GeV and 500 GeV at the ILC with the ILD detector simulation.
For the $\sqrt{s} = 250$ GeV, we assume the Higgs mass of $M_h = 120$ GeV, branching ratio of $\mathrm{Br}(h \to \tau ^+ \tau ^-) = 8.0 \ \%$, beam polarization of $P(e^-, e^+) = (-0.8, +0.3)$, and integrated luminosity of $\int L \ dt = 250 \ \mathrm{fb ^{-1}}$.
The Higgs-strahlung process $e^+ e^- \to Zh$ with $Z \to e^+ e^-$, $Z \to \mu ^+ \mu ^-$, $Z \to q\overline{q}$ mode are analyzed.
The measurement accuracy is calculated to be $\Delta (\sigma \cdot \mathrm{Br}) / (\sigma \cdot \mathrm{Br}) = 3.5 \ \%$.
The scaled result to $M_h = 125$ GeV is estimated to be $4.2 \ \%$.
For the $\sqrt{s} = 500$ GeV, we assume the Higgs mass of $M_h = 125$ GeV, branching ratio of $\mathrm{Br}(h \to \tau ^+ \tau ^-) = 6.32 \ \%$, beam polarization of $P(e^-, e^+) = (-0.8, +0.3)$, and integrated luminosity of $\int L \ dt = 500 \ \mathrm{fb ^{-1}}$.
The Higgs-strahlung process $e^+ e^- \to Zh$ with $Z \to q\overline{q}$ mode and $WW$-fusion process $e^+ e^- \to \nu _e \overline{\nu _e} h$ are analyzed.
The measurement accuracy is calculated to be $\Delta (\sigma \cdot \mathrm{Br}) / (\sigma \cdot \mathrm{Br}) = 5.7 \ \%$ for Higgs-strahlung with $Z \to q\overline{q}$ and $7.5 \ \%$ for $WW$-fusion.
\end{abstract}

\section{Introduction}

Since the discovery of Higgs boson by the ATLAS and the CMS experiments~\cite{ATLAS, CMS}, one of the next important themes for particle physics is the investigation of Higgs boson, especially the mass generation mechanism.
One of the important properties of Higgs boson is its branching ratio.
In the Standard Model (SM) of particle physics, the Yukawa coupling constant of matter fermions with the Higgs boson is proportional to the  fermion mass.
However, if there are new physics, the coupling constant will deviate from the SM prediction.
Besides, the deviation from the SM can be the few-percent level if no additional new particles are to be found at the LHC~\cite{deviate}.
Therefore, measuring the branching ratio precisely is a crucial problem from the viewpoint of new physics.

In this document, we focus on the branching ratio of $h \to \tau ^+ \tau ^-$.
We estimate the measurement accuracy $\Delta (\sigma \cdot \mathrm{Br}) / (\sigma \cdot \mathrm{Br})$ of the $h \to \tau ^+ \tau ^-$ branching ratio at $\sqrt{s} = 250$ GeV  and 500 GeV at the ILC with the ILD full detector simulation.

\section{Signal and Background}
\subsection{Signals}

\begin{figure}[!h]
\centering
\includegraphics[scale = 0.7]{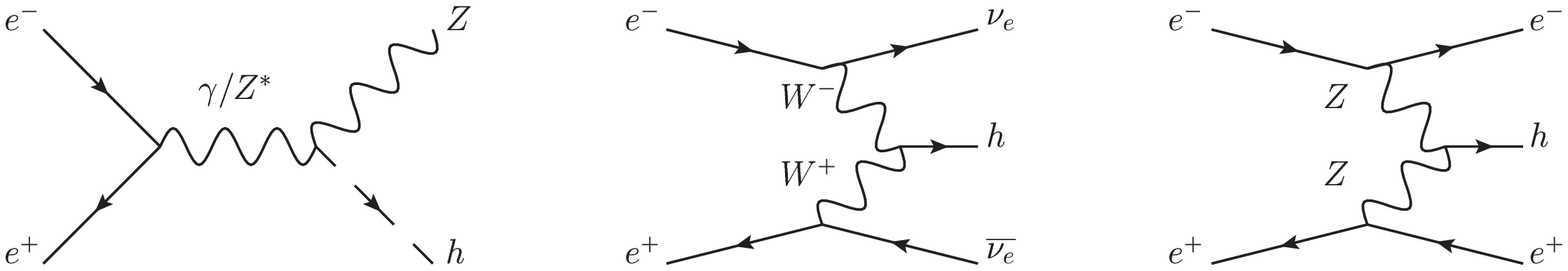}
\caption{The diagrams of Higgs production processes.
(left): Higgs-strahlung process, (middle): $WW$-fusion process, (right): $ZZ$-fusion process.}
\label{signals}
\end{figure}

There are several Higgs production process as summarized in Figure~\ref{signals}.

\begin{figure}[!h]
\centering
\includegraphics[scale = 0.65]{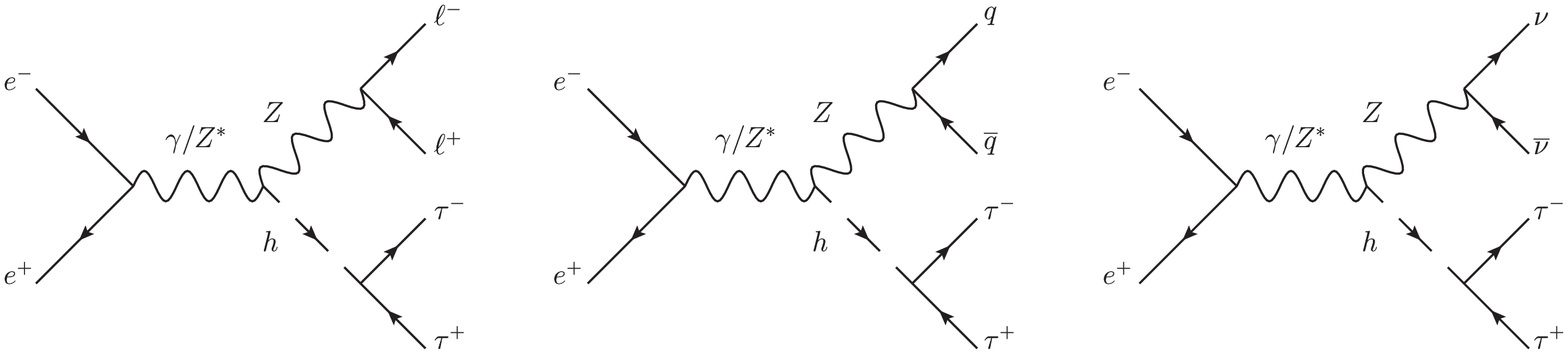}
\caption{The diagrams of Higgs-strahlung process with $Z$ boson decay.
(left): $Z \to \ell ^+ \ell ^-$ mode, (middle): $Z \to q\overline{q}$ mode, (right): $Z \to \nu \overline{\nu}$ mode.}
\label{HS}
\end{figure}

The Higgs-strahlung process $e^+ e^- \to Zh$ is the dominant process at $\sqrt{s} = 250$ GeV.
There are three types of signal depending on the decay of $Z$ boson, as shown in Figure~\ref{HS}.
The most sensitive channel at $\sqrt{s} = 250$ GeV is $Z \to q\overline{q}$ mode because of the high statistics.
In this document, we concentrate on $Z \to \ell ^+ \ell ^-$ mode and $Z \to q\overline{q}$ mode, because $Z \to \nu \overline{\nu}$ mode contributes negligibly than $Z \to q\overline{q}$ mode.
Besides, we only consider $Z \to e^+ e^-$ mode and $Z \to \mu ^+ \mu^ -$ mode as the signal process of the analysis of $Z \to \ell ^+ \ell ^-$ mode.
The cross section of $Z \to q\overline{q}$ mode is 19.8 fb, $Z \to \ell ^+ \ell ^-$ mode is 1.9 fb, respectively.

At the $\sqrt{s} = 500$ GeV, the $WW$-fusion process and $Z \to q\overline{q}$ mode of Higgs-strahlung process contributes significantly.
We concentrate on these two modes of the analyses of $\sqrt{s} = 500$ GeV, and we ignore other processes from the analyses.
The cross section of $WW$-fusion and Higgs-strahlung at $\sqrt{s} = 500$ GeV is 149.5 fb and 100.4 fb.

\subsection{Backgrounds}

\begin{figure}[!h]
\centering
\includegraphics[scale = 0.7]{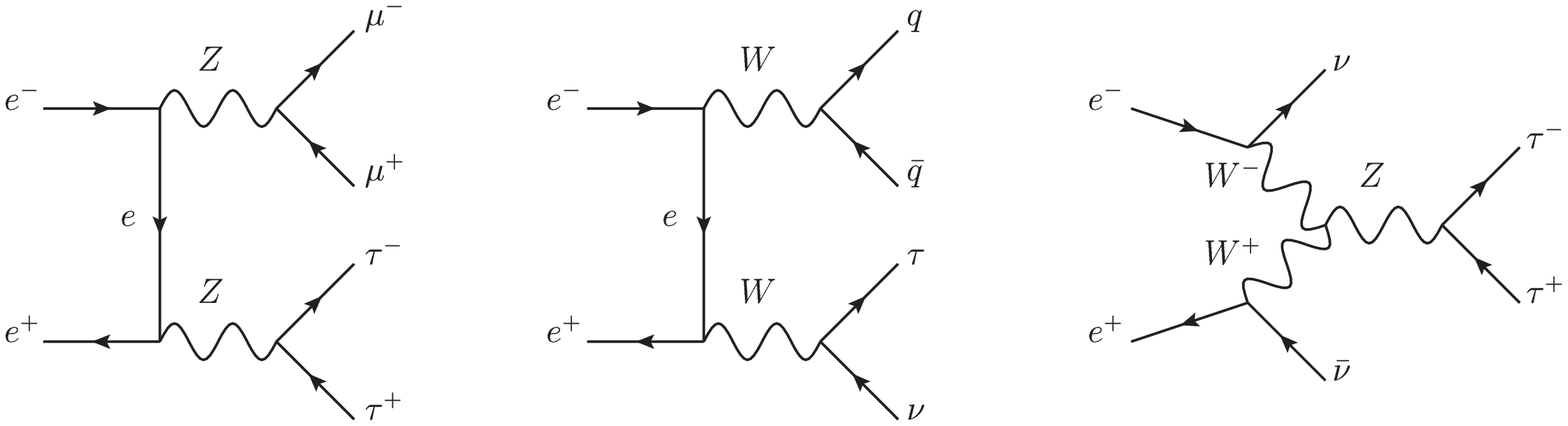}
\caption{The example diagrams of backgrounds.
(left): $\mu ^+ \mu ^- \tau ^+ \tau ^-$ background for $Z\to \ell ^+ \ell ^-$ mode, (middle): $q\overline{q}\tau \nu$ background for $Z \to q\overline{q}$ mode, (right): $\nu \overline{\nu} \tau ^+ \tau ^-$ background for $WW$-fusion process.}
\label{backgrounds}
\end{figure}

For the $Z \to \ell ^+ \ell ^-$ mode, the possible backgrounds are the processes which including four leptons in the final state.
The left diagram of Figure~\ref{backgrounds} shows the example of $\mu ^+ \mu ^- \tau ^+ \tau ^-$ process via $e^+ e^- \to ZZ$.
Other possible processes are $e^+ e^- \to Zh$ reactions with the Higgs boson does not decay to tau pairs ($H \not \to \tau ^+ \tau ^-$).

For the $Z \to q \overline{q}$ mode, the $q \overline{q} q \overline{q}$, $q \overline{q} \ell ^+ \ell ^-$, and $q \overline{q} \ell \nu$ which comes from $e^+ e^- \to W^+ W^-$ and $e^+ e^- \to ZZ$ processes should be the main background.
The middle diagram of Figure~\ref{backgrounds} shows the background of $e^+ e^- \to WW \to q\overline{q} \tau \nu$ process.

On the other hand, the possible backgrounds for $WW$-fusion process are $e^+e^- \to W^+ W^- \to \tau ^+ \overline{\nu} \tau ^- \nu$ and $e^+ e^- \to \nu \nu Z$ with $Z \to \tau ^+ \tau ^-$.
The diagram of latter process is shown in the right of Figure~\ref{backgrounds}.

\section{Simulation Conditions}

We perform the detector simulation with \verb|Mokka|~\cite{Mokka}, a Geant4-based~\cite{Geant4} full simulation, with the ILD detector model.
\verb|TAUOLA|~\cite{TAUOLA} is used for the tau decay simulation.
The ILD detector model is consists of a vertex detector, a time projection chamber (TPC), an electromagnetic calorimeter (ECAL), a hadronic calorimeter (HCAL), a return yoke, muon systems, and forward components.

For the analysis of $\sqrt{s} = 250$ GeV, we use the signal and background samples which were generated in the context of the Letter of Intent~\cite{ILDLOI}, and use \verb|ILD_00| detector model.
The effects of beamstrahlung and initial state radiation are included.
We assume a Higgs mass of $M_h = 120$ GeV, a branching ratio of $\mathrm{Br}(h \to \tau ^+ \tau ^-) = 8.0 \ \%$ as assumed by \verb|PYTHIA|~\cite{PYTHIA}, an integrated luminosity of $\int L \ dt = 250 \ \mathrm{fb^{-1}}$, and a beam polarization of $P(e^+, e^-) = (+0.3, -0.8)$.
We also rescale the final result to the case of $M_h = 125$ GeV and the $h \to \tau ^+ \tau ^-$ branching ratio which includes the NNLO corrections~\cite{NNLO}.

For the analysis of $\sqrt{s} = 500$ GeV, we use the signal and background samples which were generated in the context of ILC Technical Design Report~\cite{TDR1, TDR2, TDR3, TDR4}, and use \verb|ILD_o1_v05| model.
In these samples, the effects of $\gamma \gamma \to$ hadron(s) overlay process are also included.
We use the processes of $q\overline{q}h$, $\nu \overline{\nu} h$, $\ell ^+ \ell ^- h$, 2f, 4f, 5f, 6f, and $\gamma \gamma \to$ 4f (f $=$ fermions).
We assume a Higgs mass of $M_h = 125$ GeV, a branching ratio of $\mathrm{Br}(h \to \tau ^+ \tau ^-) = 6.32 \ \%$~\cite{NNLO}, an integrated luminosity of $\int L \ dt = 500 \ \mathrm{fb^{-1}}$, and a beam polarization of $P(e^+, e^-) = (+0.3, -0.8)$.

\section{Event Reconstruction and Event Selection}

\subsection{$Z \to \ell ^+ \ell ^-$ mode at $\sqrt{s} = 250$ GeV}

In this mode, we take the strategy of reconstructing the $Z$ boson first, followed by the reconstruction of the tau pairs from the Higgs decay.

\begin{figure}[!h]
\begin{tabular}{cc}
\begin{minipage}{0.47\textwidth}
\centering
\includegraphics[width = 8.0truecm]{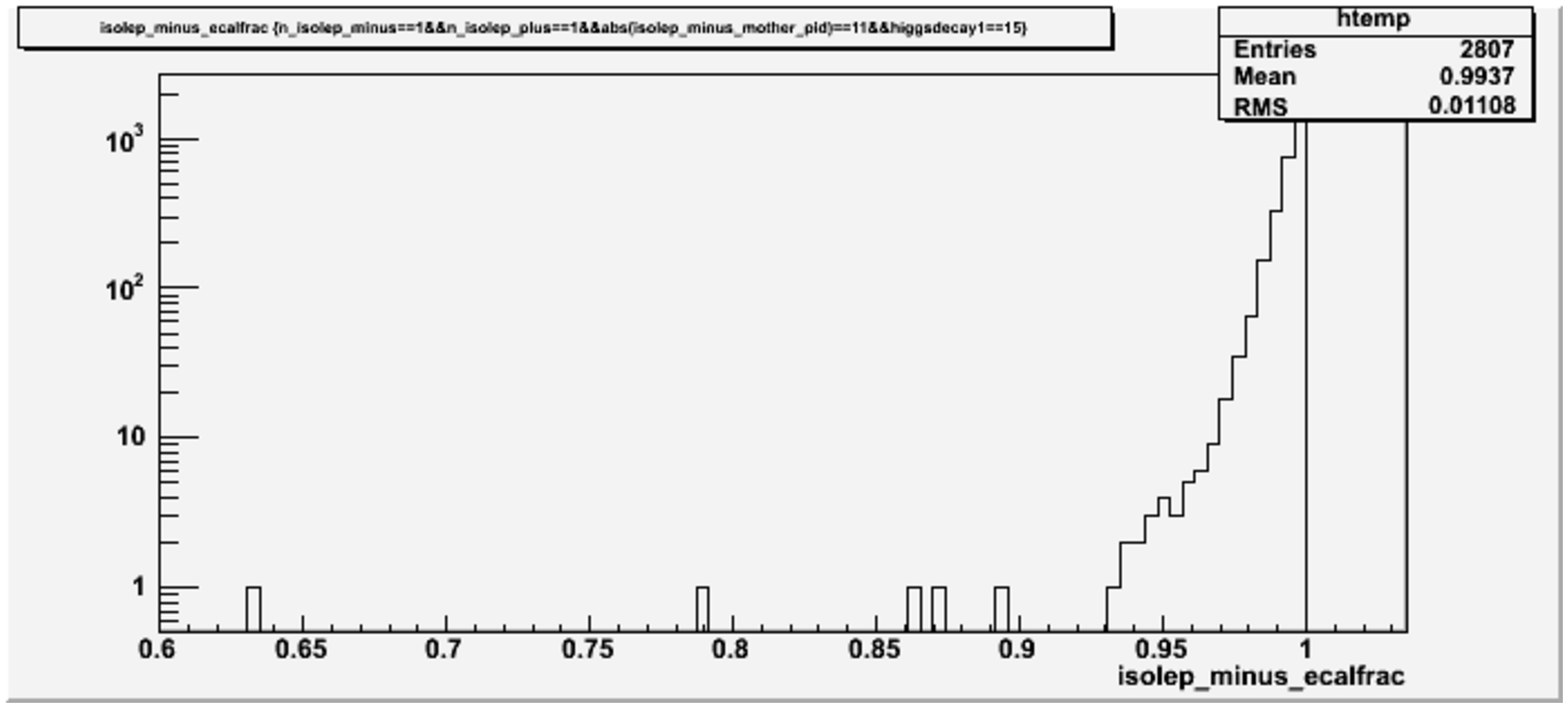}
\caption{The plot of $E_{\mathrm{ECAL}} / (E_{\mathrm{ECAL}} + E_{\mathrm{HCAL}})$ for the $e$ in $e^+ e^- h$ samples.}
\label{ecalfrac_e1}
\end{minipage}
\begin{minipage}{0.47\textwidth}
\centering
\includegraphics[width = 8.0truecm]{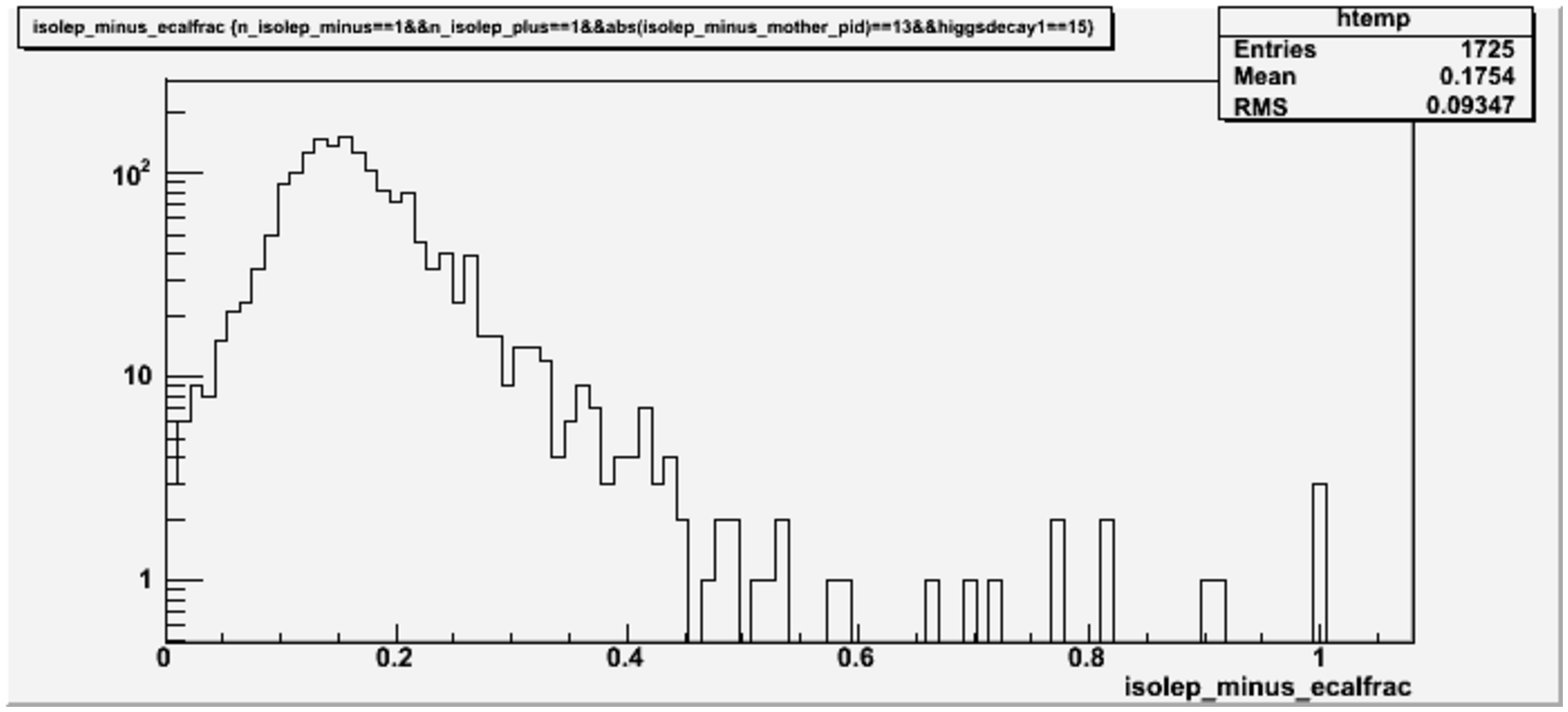}
\caption{The plot of $E_{\mathrm{ECAL}} / (E_{\mathrm{ECAL}} + E_{\mathrm{HCAL}})$ for the $\mu$ in $\mu ^+ \mu ^- h$ samples.}
\label{ecalfrac_e2}
\end{minipage}
\end{tabular}
\end{figure}

\begin{figure}[!h]
\begin{tabular}{cc}
\begin{minipage}{0.47\textwidth}
\centering
\includegraphics[width = 8.0truecm]{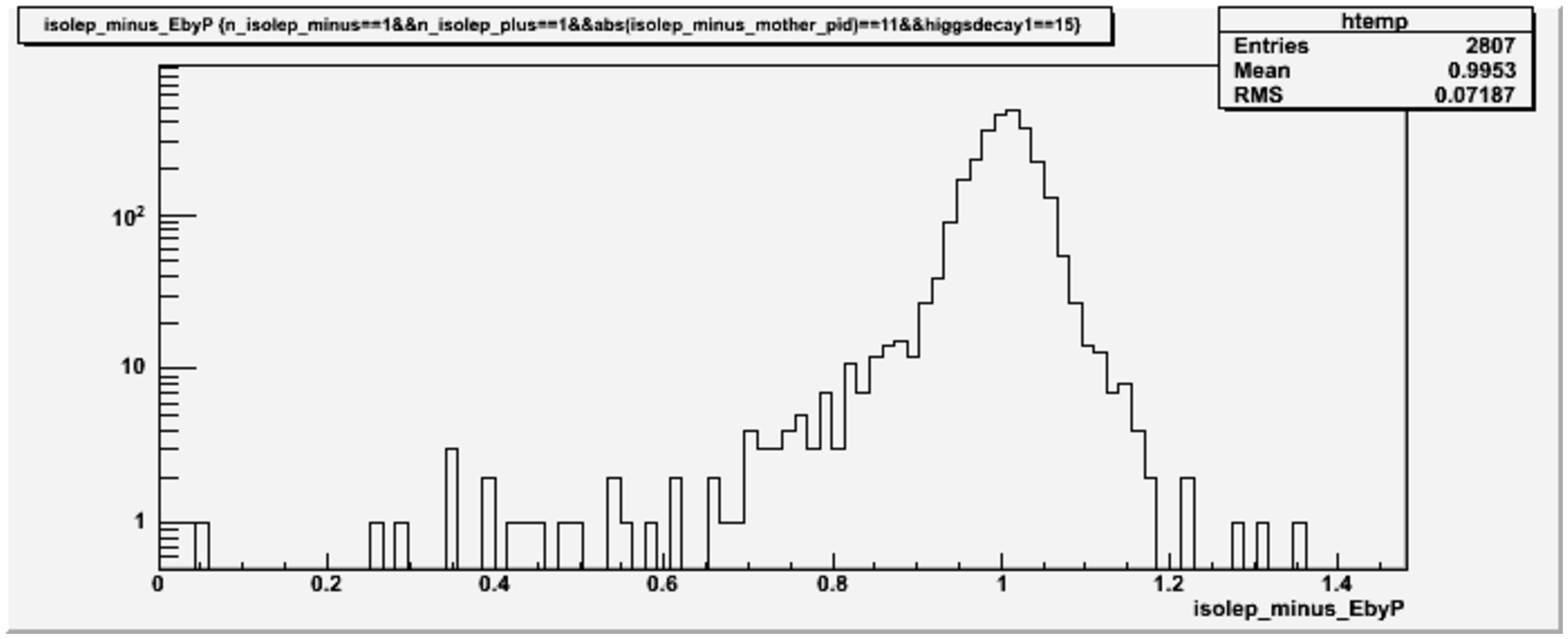}
\caption{The plot of $(E_{\mathrm{ECAL}} + E_{\mathrm{HCAL}}) / P_{\mathrm{track}}$ for the $e$ in $e^+ e^-h$ samples.}
\label{EbyP_e1}
\end{minipage}
\begin{minipage}{0.47\textwidth}
\centering
\includegraphics[width = 8.0truecm]{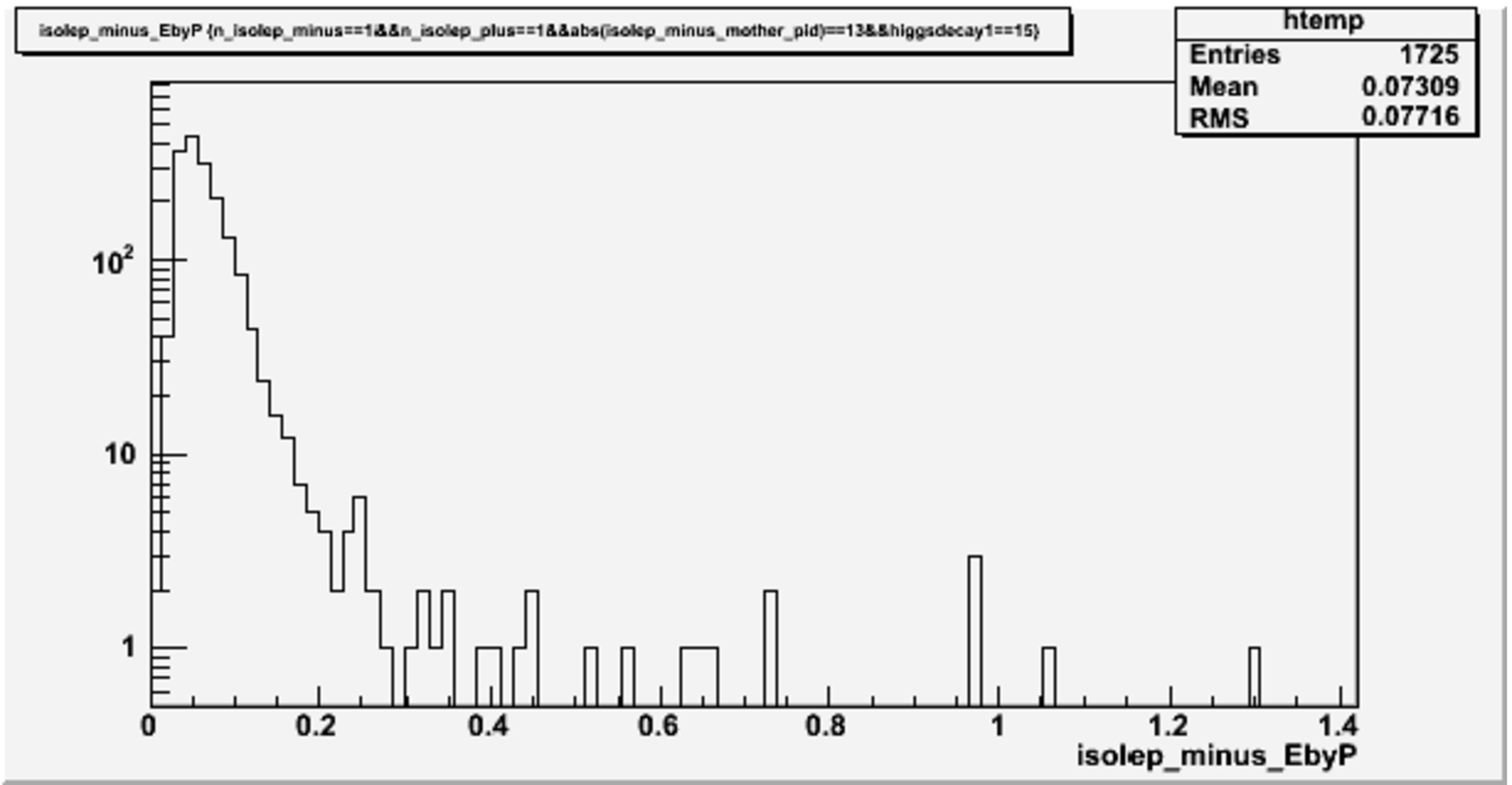}
\caption{The plot of $(E_{\mathrm{ECAL}} + E_{\mathrm{HCAL}}) / P_{\mathrm{track}}$ for the $\mu$ in $\mu ^+ \mu ^- h$ samples.}
\label{EbyP_e2}
\end{minipage}
\end{tabular}
\end{figure}

We apply lepton identification at first for dividing $Z \to e^+ e^-$ events and $Z \to \mu ^+ \mu ^-$ events by using the information of energy deposit in the calorimeter ($E_{\mathrm{ECAL}}$ and $E_{\mathrm{HCAL}}$, where $E_{\mathrm{ECAL(HCAL)}}$ is the energy deposit in ECAL(HCAL)) and track momentum ($P_{\mathrm{track}}$).
Figures~\ref{ecalfrac_e1} -~\ref{EbyP_e2} are the plots of $E_{\mathrm{ECAL}} / (E_{\mathrm{ECAL}} + E_{\mathrm{HCAL}})$ and $(E_{\mathrm{ECAL}} + E_{\mathrm{HCAL}}) / P_{\mathrm{track}}$.

From these plots, we define the criteria for lepton identification.
The criteria for electron identification ($e$-ID) and muon identification ($\mu$-ID) are summarized in Table~\ref{tab:Ztoll250_ID}.

\begin{table}[!h]
\centering
\caption{The criteria for lepton identification for $Z \to \ell ^+ \ell ^-$ mode at $\sqrt{s} = 250$ GeV.}
\begin{tabular}{c|c|c} \hline
 & $e$-ID & $\mu$-ID \\ \hline
$E_{\mathrm{ECAL}} / (E_{\mathrm{ECAL}} + E_{\mathrm{HCAL}})$ & $> 0.92$ & $< 0.6$ \\
$(E_{\mathrm{ECAL}} + E_{\mathrm{HCAL}}) / P_{\mathrm{track}}$ & $> 0.5$ & $< 0.5$ \\ \hline
\end{tabular}
\label{tab:Ztoll250_ID}
\end{table}

\begin{figure}[!h]
\begin{tabular}{cc}
\begin{minipage}{0.47\textwidth}
\centering
\includegraphics[width = 8.0truecm]{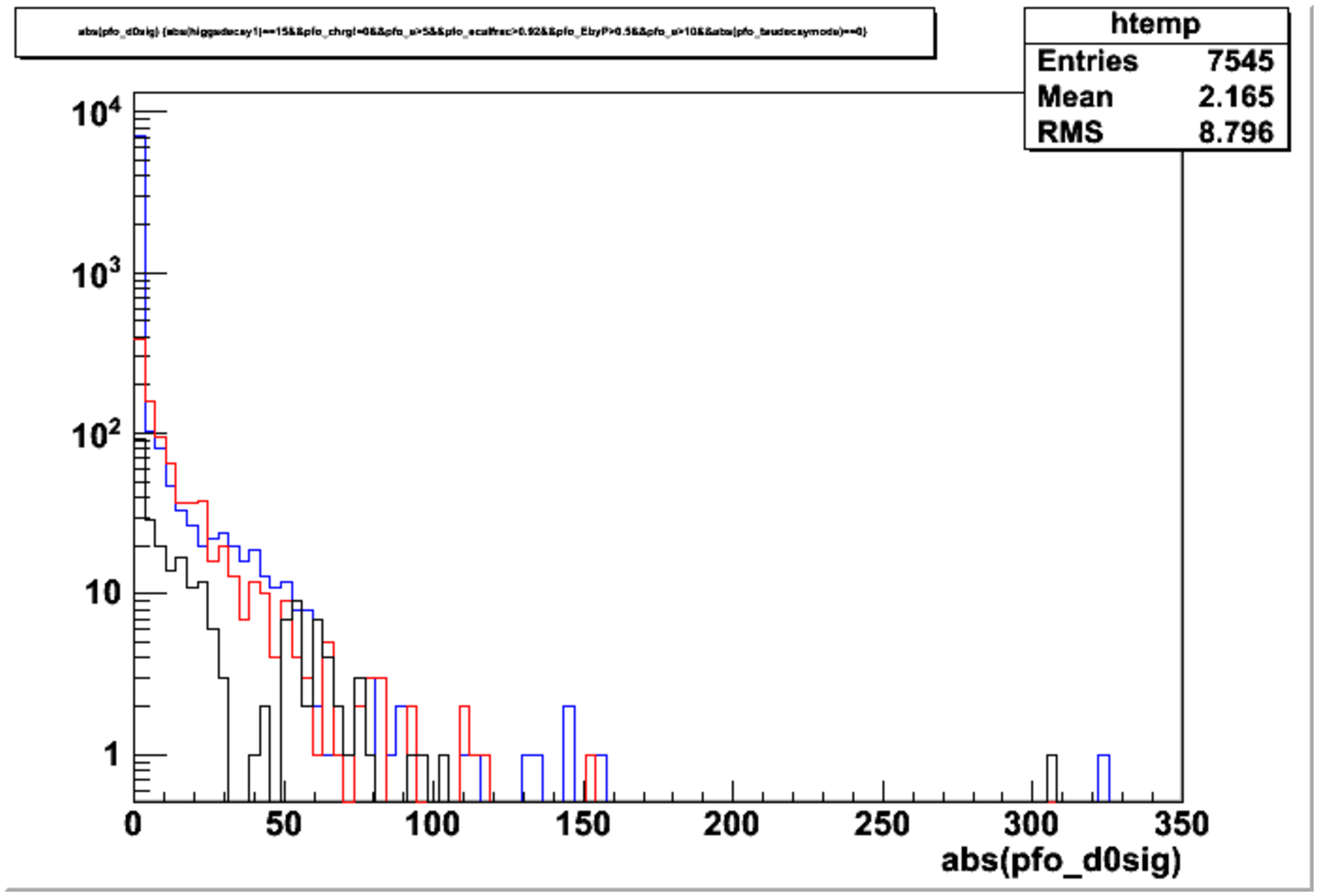}
\caption{The plot of $|d_0 / \sigma(d_0)|$ of $e$ of $e^+ e^- h$ process.
Blue, red, and black histograms show the $e$ from $Z \to e^+ e^-$, the $e$ from $\tau \to e \nu _{\tau} \nu _e$, and the hadrons from $\tau$ decay, respectively.}
\label{taureject_EID_d0sig}
\end{minipage}
\begin{minipage}{0.47\textwidth}
\centering
\includegraphics[width = 8.0truecm]{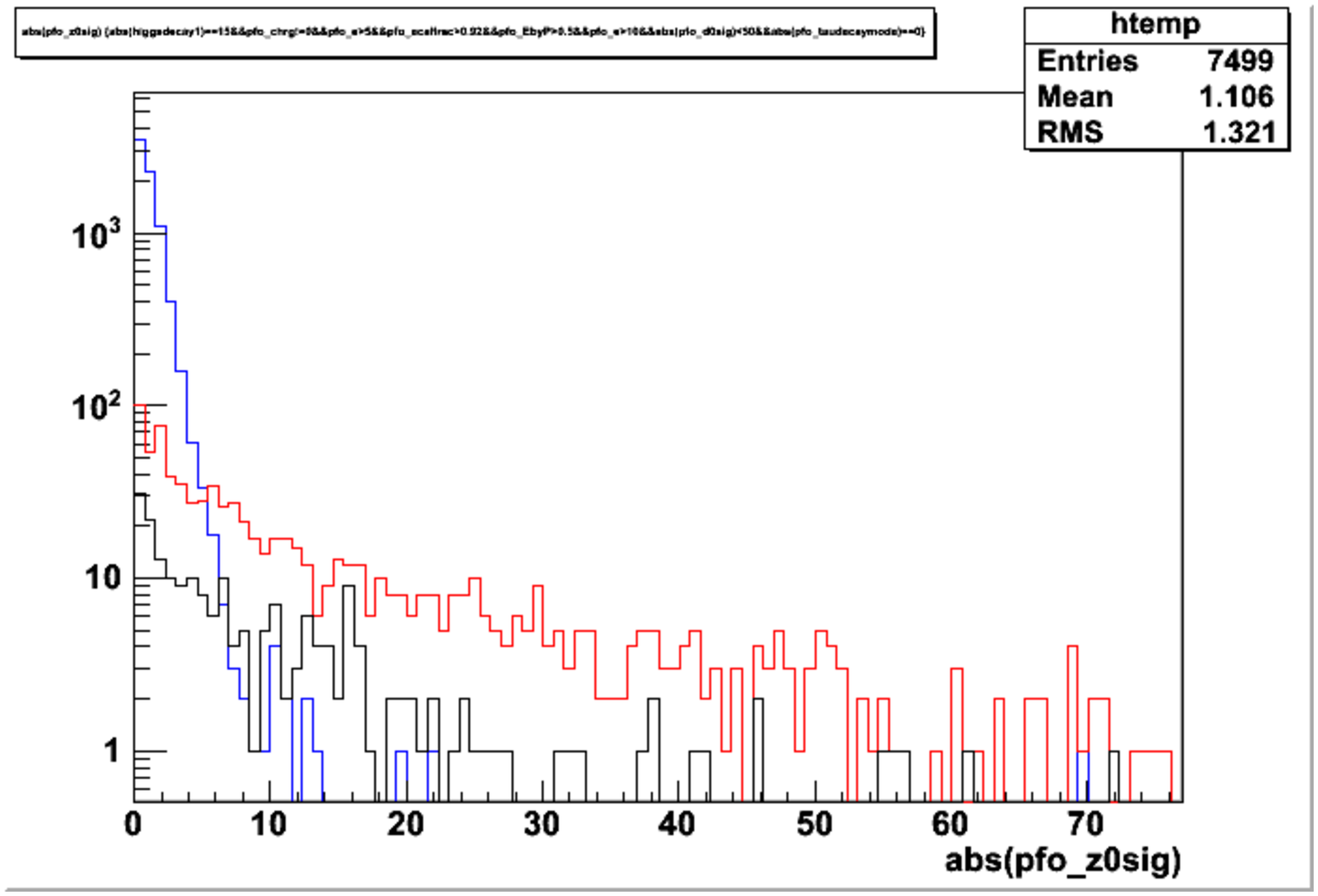}
\caption{The plot of $|z_0 / \sigma(z_0)|$ of $e$ of $e^+ e^- h$ process.
Blue, red, and black histograms show the $e$ from $Z \to e^+ e^-$, the $e$ from $\tau \to e \nu _{\tau} \nu _e$, and the hadrons from $\tau$ decay, respectively.}
\label{taureject_EID_z0sig}
\end{minipage}
\end{tabular}
\end{figure}

\begin{figure}[!h]
\begin{tabular}{cc}
\begin{minipage}{0.47\textwidth}
\centering
\includegraphics[width = 8.0truecm]{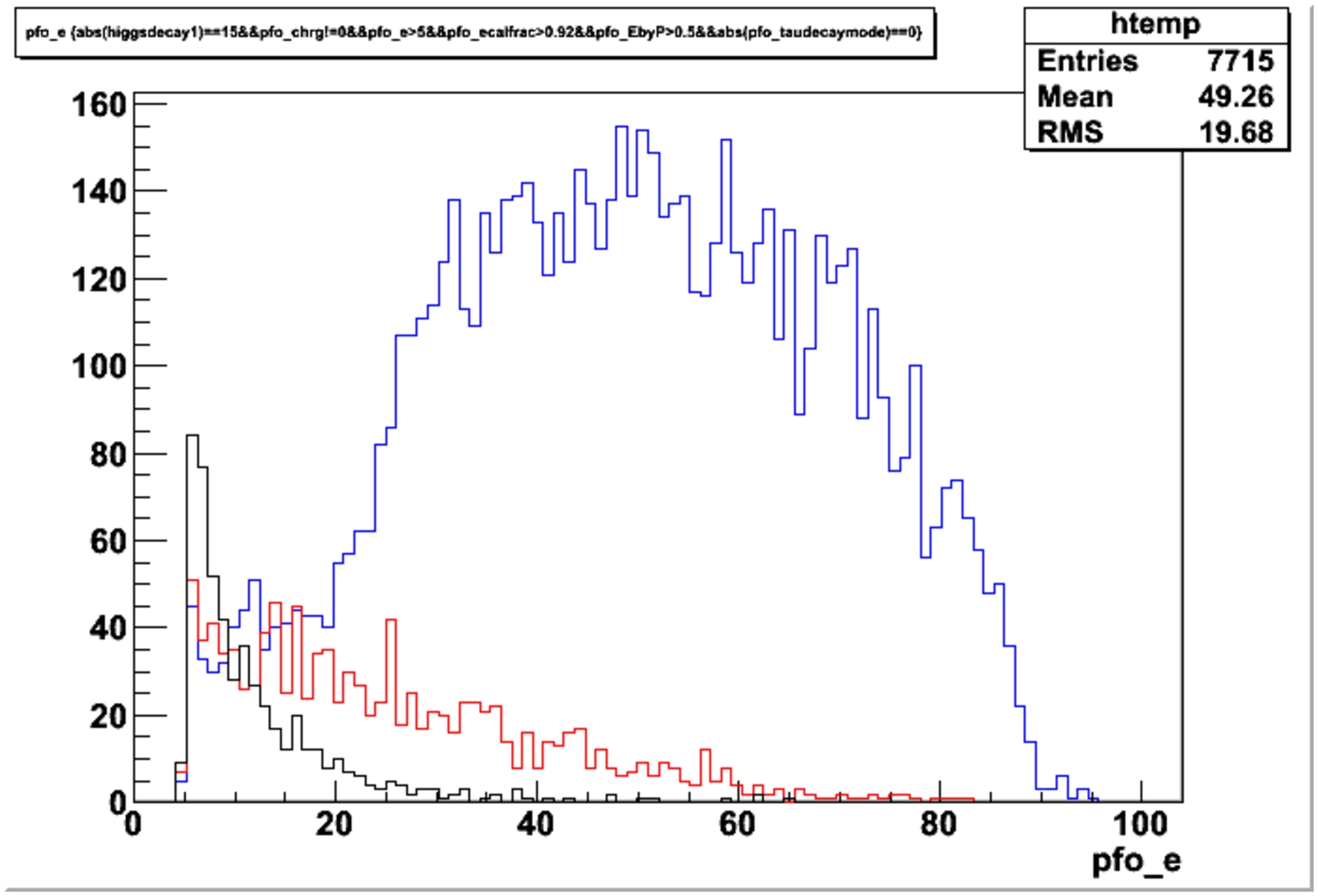}
\caption{The plot of $E_{\mathrm{track}}$ of $e$ of $e^+ e^- h$ process.
Blue, red, and black histograms show the $e$ from $Z \to e^+ e^-$, the $e$ from $\tau \to e \nu _{\tau} \nu _e$, and the hadrons from $\tau$ decay, respectively.}
\label{taureject_EID_E}
\end{minipage}
\begin{minipage}{0.47\textwidth}
\centering
\includegraphics[width = 8.0truecm]{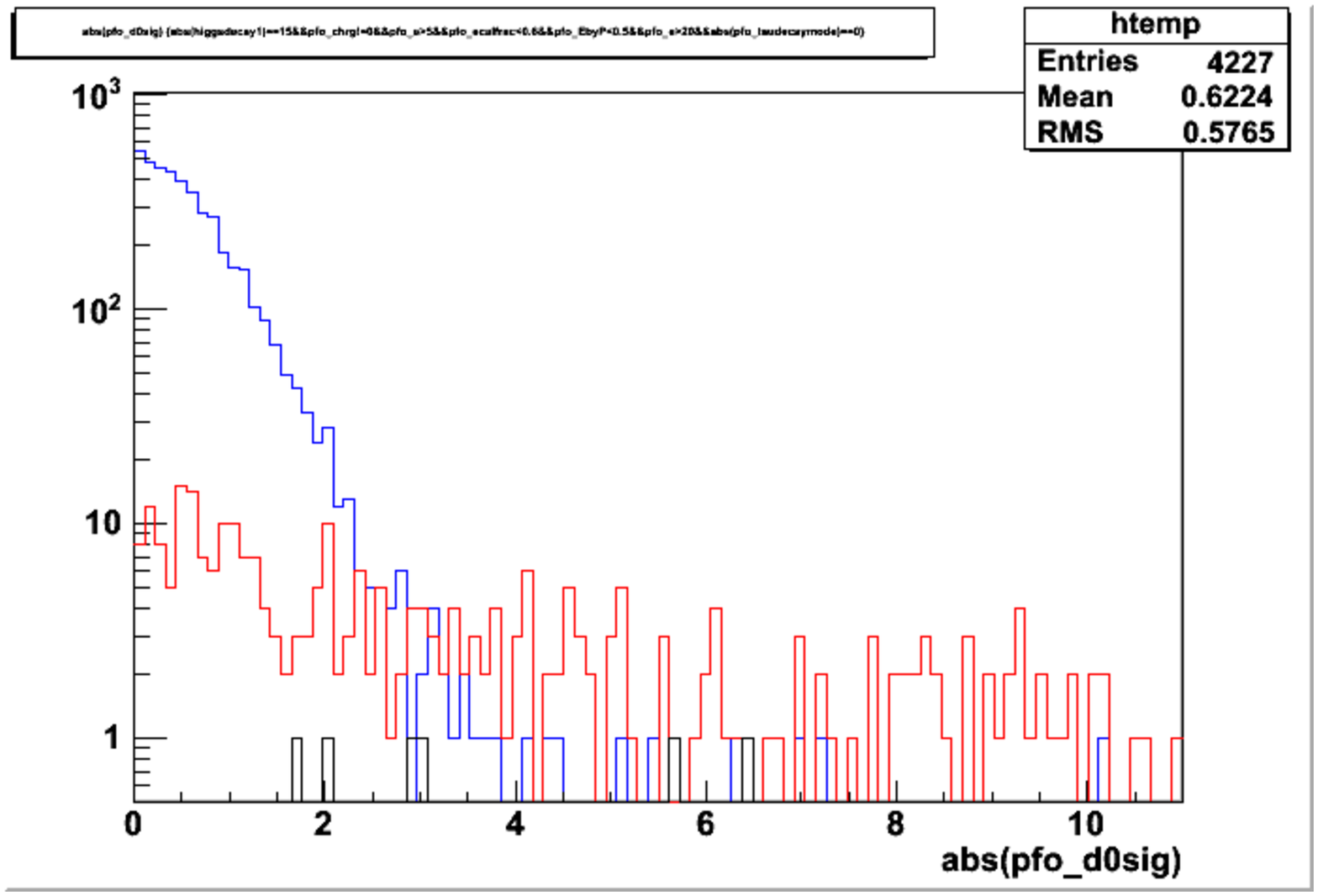}
\caption{The plot of $|d_0 / \sigma(d_0)|$ of $\mu$ of $\mu ^+ \mu ^- h$ process.
Blue, red, and black histograms show the $\mu$ from $Z \to \mu ^+ \mu ^-$, the $\mu$ from $\tau \to \mu \nu _{\tau} \nu _{\mu}$, and the hadrons from $\tau$ decay, respectively.}
\label{taureject_MuID_d0sig}
\end{minipage}
\end{tabular}
\end{figure}

\begin{figure}[!h]
\begin{tabular}{cc}
\begin{minipage}{0.47\textwidth}
\centering
\includegraphics[width = 8.0truecm]{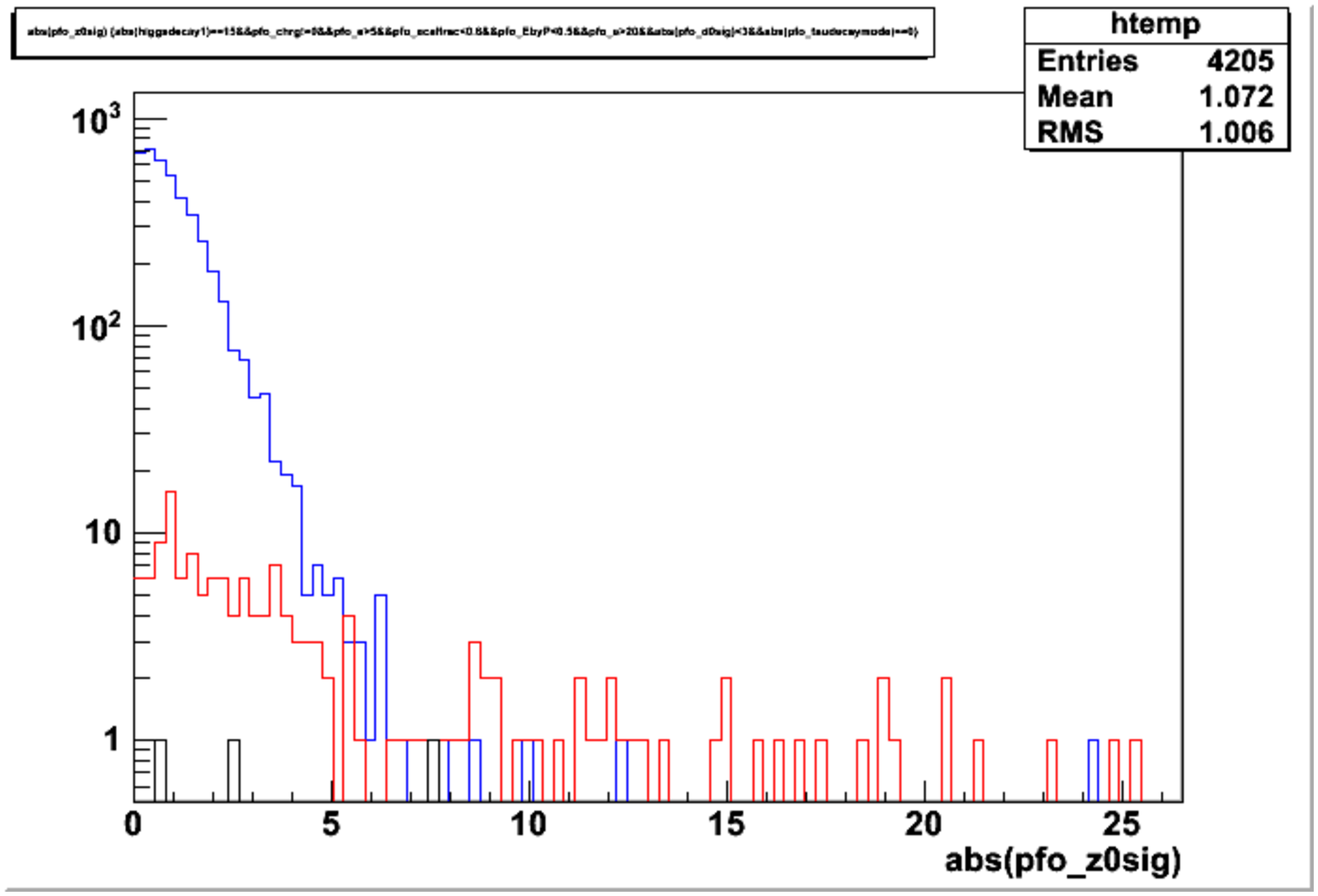}
\caption{The plot of $|z_0 / \sigma(z_0)|$ of $\mu$ of $\mu ^+ \mu ^- h$ process.
Blue, red, and black histograms show the $\mu$ from $Z \to \mu ^+ \mu ^-$, the $\mu$ from $\tau \to \mu \nu _{\tau} \nu _{\mu}$, and the hadrons from $\tau$ decay, respectively.}
\label{taureject_MuID_z0sig}
\end{minipage}
\begin{minipage}{0.47\textwidth}
\centering
\includegraphics[width = 8.0truecm]{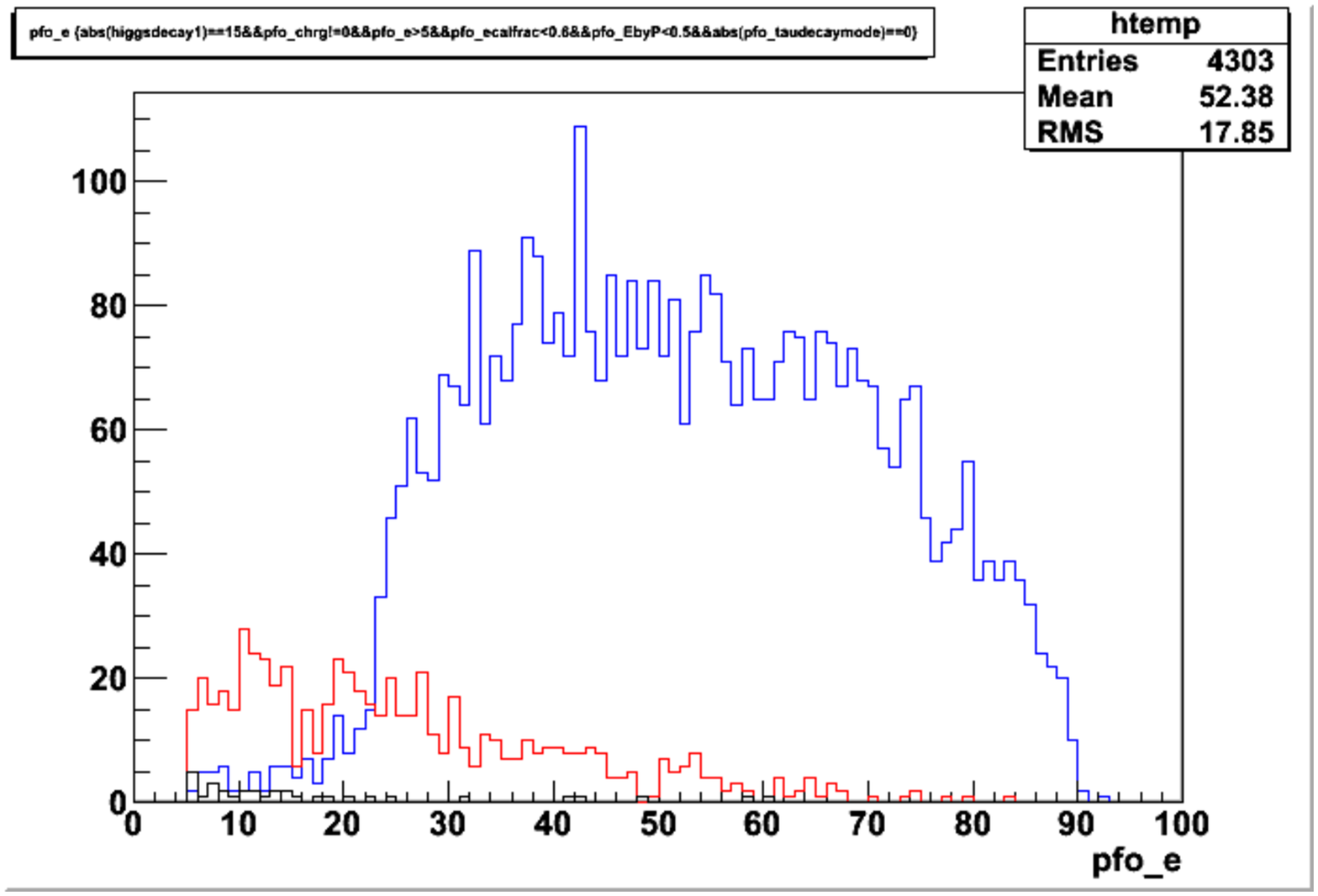}
\caption{The plot of $E_{\mathrm{track}}$ of $\mu$ of $\mu ^+ \mu ^- h$ process.
Blue, red, and black histograms show the $\mu$ from $Z \to \mu ^+ \mu ^-$, the $\mu$ from $\tau \to \mu \nu _{\tau} \nu _{\mu}$, and the hadrons from $\tau$ decay, respectively.}
\label{taureject_MuID_E}
\end{minipage}
\end{tabular}
\end{figure}

After the lepton identification, we apply selections to remove secondary leptons from tau decays.
The strategy of this selection is to remove tracks which do not come from the interaction point (IP) by using the track energy $E_{\mathrm{track}}$ and impact parameter in the transverse direction $d_0$ and longitudinal direction $z_0$ with respect to the beam axis.
Figures~\ref{taureject_EID_d0sig} -~\ref{taureject_MuID_E} show the $|d_0 / \sigma(d_0)|$, $|z_0 / \sigma(z_0)|$, and $E_{\mathrm{track}}$ plots which through the lepton identification.

We define the tau rejection cut for the objects through the $e$-ID and the $\mu$-ID respectively as shown in Table~\ref{tab:Ztoll250_taureject}.

\begin{table}[!h]
\centering
\caption{The criteria for lepton identification for $Z \to \ell ^+ \ell ^-$ mode at $\sqrt{s} = 250$ GeV.}
\begin{tabular}{c|c|c} \hline
 & $e$-ID & $\mu$-ID \\ \hline
$|d_0 / \sigma (d_0)|$ & $< 50$ & $< 3$ \\
$|z_0 / \sigma (z_0)|$ & $< 5$ & $< 7$ \\
$E_{\mathrm{track}}$ & $> 10$ GeV & $> 20$ GeV \\ \hline
\end{tabular}
\label{tab:Ztoll250_taureject}
\end{table}

\begin{figure}[!h]
\centering
\includegraphics[width = 5.0truecm]{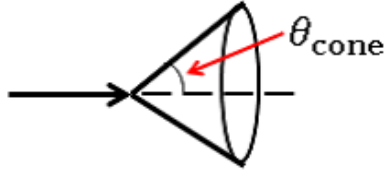}
\caption{The definition of the cone. 
Black arrow shows the lepton candidate.
$\theta _{\mathrm{cone}}$ is the angle of the cone.}
\vspace*{-\intextsep}
\label{cone}
\end{figure}

\begin{figure}[!h]
\begin{tabular}{cc}
\begin{minipage}{0.47\textwidth}
\centering
\includegraphics[width = 8.0truecm]{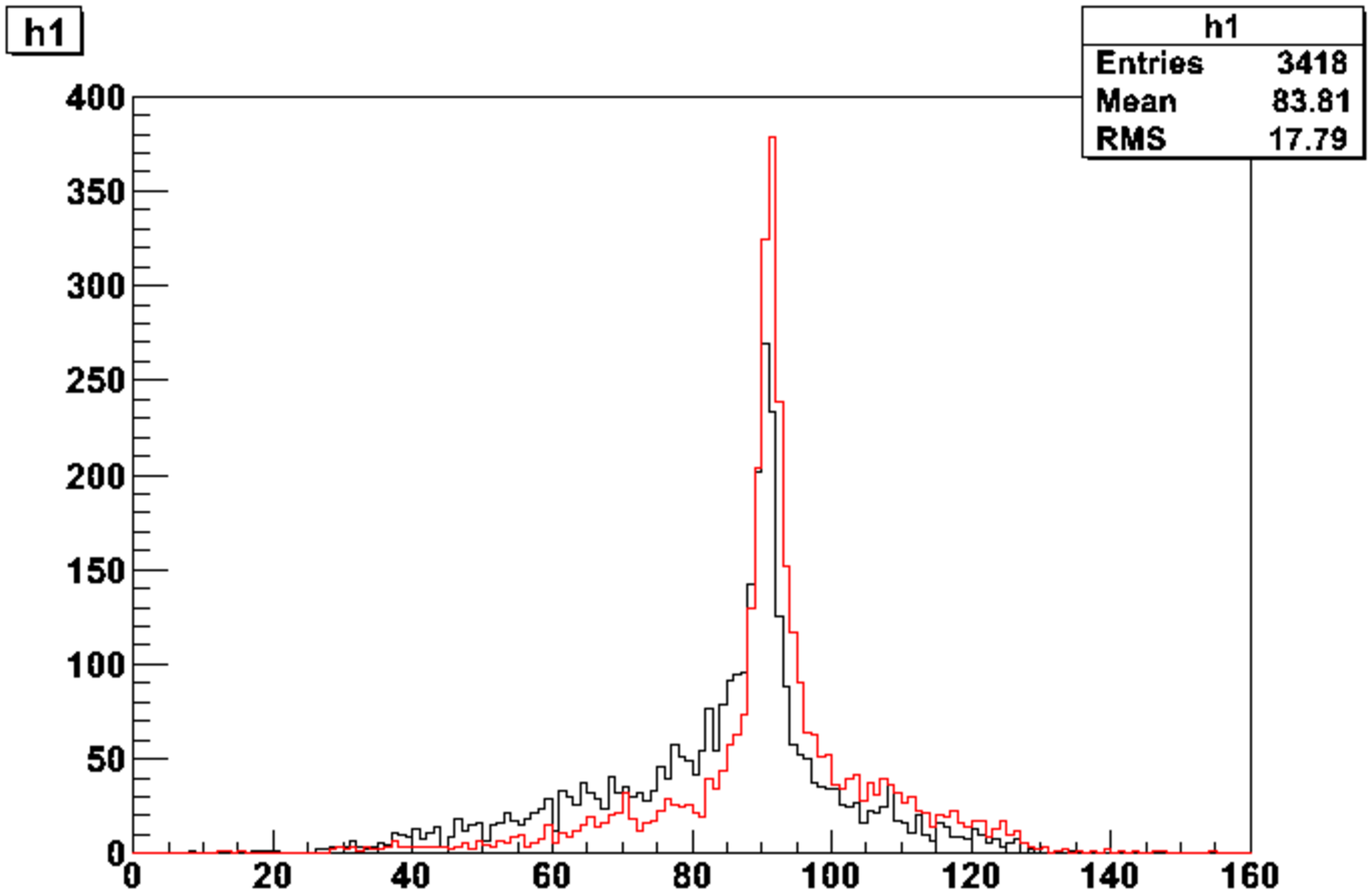}
\caption{The results of recovery for $Z \to e^+ e^-$ mode.
The horizontal axis shows the $M_Z$.
Black and red histograms show the results of without recovery and with recovery ($\cos \theta _{\mathrm{cone}} = 0.999$), respectively.}
\label{recovery_e1e1h}
\end{minipage}
\begin{minipage}{0.47\textwidth}
\centering
\includegraphics[width = 8.0truecm]{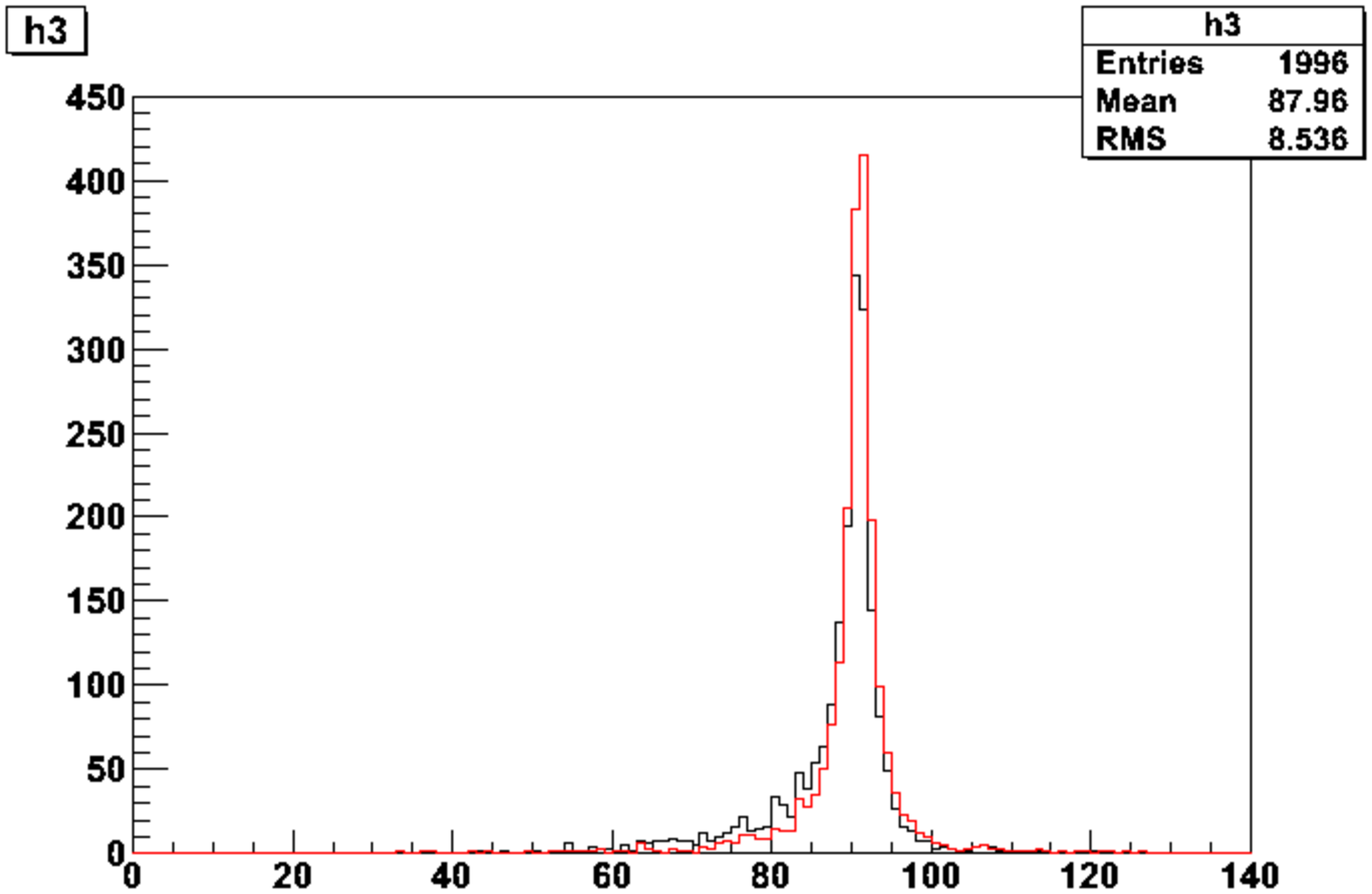}
\caption{The results of recovery for $Z \to \mu ^+ \mu ^-$ mode.
The horizontal axis shows the $M_Z$.
Black and red histograms show the results of without recovery and with recovery ($\cos \theta _{\mathrm{cone}} = 0.999$), respectively.}
\label{recovery_e2e2h}
\end{minipage}
\end{tabular}
\end{figure}

We apply the energy recovery procedure to correct the effect of bremsstrahlung and final state radiation.
In order to reconstruct the original $Z$ boson, we have to use both the charged particles and the radiated photons.
To achieve this, we define the cone as shown in Figure~\ref{cone}.
The four-momenta of the neutral particles in the cone are combined with that of the lepton candidate.
We define the half-opening angle of the cone with $\cos \theta _{\mathrm{cone}} = 0.999$ and apply the recovery procedure to the lepton candidates.
The results are shown in Figures~\ref{recovery_e1e1h} and \ref{recovery_e2e2h}.

After that, we apply the tau finder to the remaining objects to reconstruct tau leptons.
First of all, the objects which already used at $Z$ boson reconstruction are rejected from tau reconstruction analysis.
Then the finder searches the highest energy track from the remaining objects, and combine the neighboring particles (which satisfies the angle with respect to the highest energy track less than 1.0 radian) with the combined mass less than 2 GeV.
We regard the combined object as a tau candidate.
Then repeat these processes until there are no charged particles.

After finishing the event reconstruction, we apply the cuts for selecting signal, rejecting background.
Before optimizing the cuts, we apply the preselection as follows for $Z \to e^+ e^-$ mode: number of $e^+$ and $e^- = 1$, number of $\tau ^+$ and $\tau ^- = 1$, and for $Z \to \mu ^+ \mu ^-$ mode: number of $\mu ^+$ and $\mu ^- = 1$, number of  $\tau ^+$ and $\tau ^- = 1$.

\begin{figure}[!h]
\centering
\includegraphics[width = 10.0truecm]{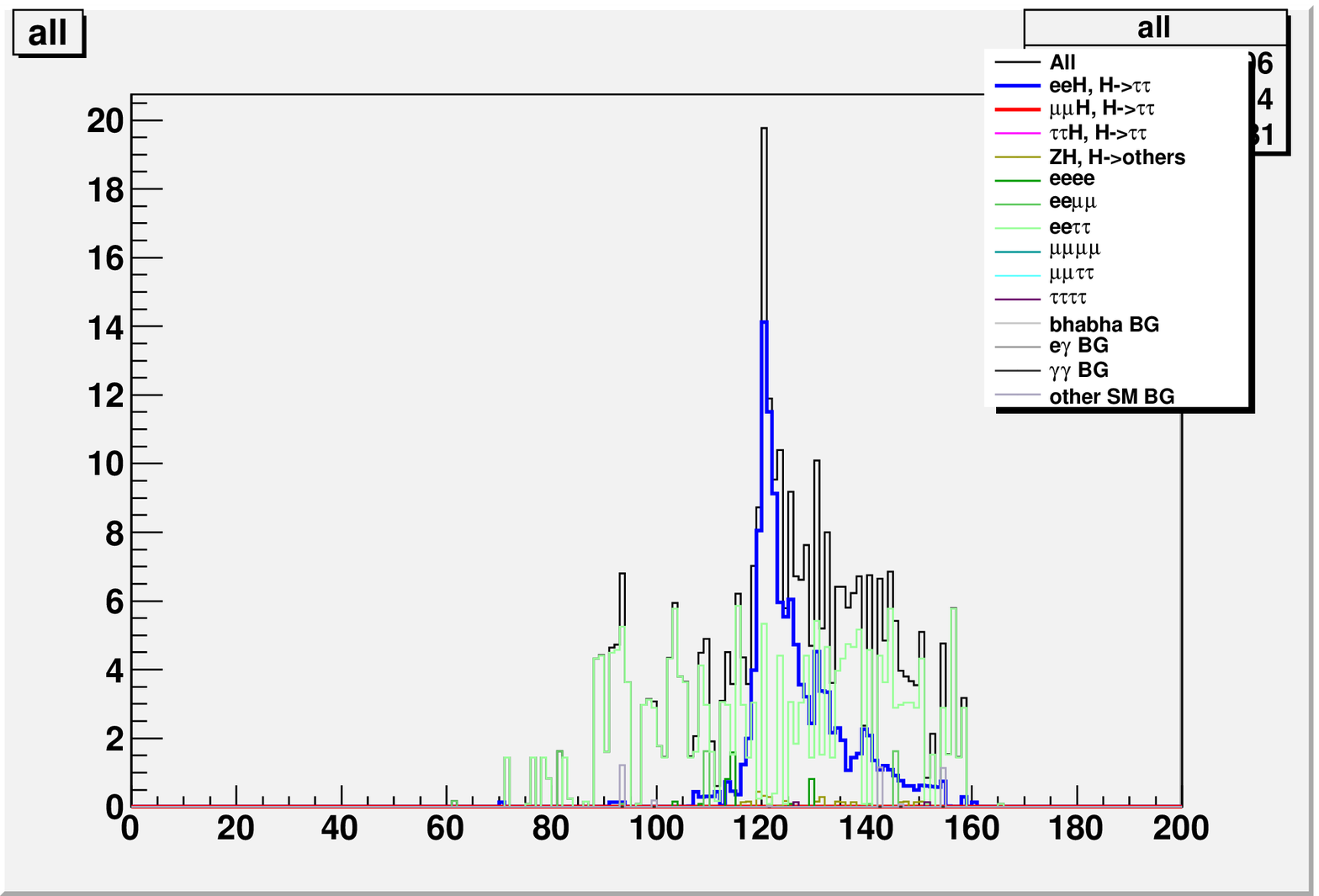}
\caption{The distribution of recoil mass in the unit of GeV.}
\label{Ztoee250_recoilmass}
\end{figure}

We apply the following cuts sequentially for $Z \to e^+ e^-$ mode: number of tracks $\le 8$, 115 GeV $< E_{\mathrm{vis}} < 230$ GeV, $|\cos \theta _{\mathrm{miss}}| < 0.99$, 81 GeV $< M_Z < 113$ GeV, $\cos \theta _{e^-} < 0.92$, $\cos \theta _{e^+} > -0.92$, $E_{e^- (e^+)} < 90$ GeV, $\cos \theta _{\tau ^+ \tau ^-} < -0.45$, $\cos \theta _{\tau ^-} < 0.92$, $\cos \theta _{\tau ^+} > -0.92$, and 116 GeV $< M_{\mathrm{recoil}} < 142$ GeV, where $E_{\mathrm{vis}}$ is the visible energy, $\theta _{\mathrm{miss}}$ is the missing momentum angle with respect to beam axis, $\theta _{e^-(e^+)}$ is the $e^-(e^+)$ angle with respect to beam axis, $E_{e^-(e^+)}$ is the $e^-(e^+)$ energy, $\theta _{\tau ^+ \tau ^-}$ is the angle between $\tau ^+$ and $\tau ^-$, $\theta _{\tau ^-(\tau ^+)}$ is the $\tau ^-(\tau ^+)$ angle with respect to beam axis, and $M_{\mathrm{recoil}}$ is the recoil mass, respectively.
Figure~\ref{Ztoee250_recoilmass} shows the recoil mass distribution.
Table~\ref{tab:Ztoee250_cuttable} shows the cut statistics of this mode.
After the cuts, the $Z \to e^+ e^-$ signal events of 108.9 and background events of 76.0 remained.
The statistical significance is calculated to be $S / \sqrt{S + B} = 108.9 / \sqrt{108.9 + 76.0} = 8.0 \sigma$.

\begin{table}[!h]
\centering
\caption{The cut statistics of $Z \to e^+ e^-$ mode at $\sqrt{s} = 250$ GeV.}
\begin{tabular}{c|cccccccc}\hline
\rule[-1pt]{0pt}{10pt} & $e^+ e^- h$ & $\mu ^+ \mu ^- h$ & $\tau ^+ \tau ^- h$ & $Zh$ with & $e^+ e^- \tau ^+ \tau ^-$ & other & other  & signi. \\
 & $h \to \tau ^+ \tau ^-$ & $h \to \tau ^+ \tau ^-$ & $h \to \tau ^+ \tau ^-$ & $h \not \to \tau ^+ \tau ^-$ &  & 4 leptons & SM bkg. & \\ \hline
\rule[-1pt]{0pt}{10pt} No cut & 228.3 & 211.1 & 214.6 & 7325 & $2.388 \times 10^5$ & $5.238 \times 10^5$ & $1.492 \times 10^{10}$ & 0.0019 \\
preselection & 171.3 & 0.155 & 1.532 & 47.05 & $1.338 \times 10^4$ & $3.215 \times 10^4$ & $1.023 \times 10^7$ & 0.053 \\
\# of tracks & 169.4 & 0.155 & 1.532 & 41.56 & $1.316 \times 10^4$ & $3.205 \times 10^4$ & $1.009 \times 10^7$ & 0.053 \\
$E_{\mathrm{vis}}$ & 162.3 & 0.155 & 0.912 & 38.36 & $1.068 \times 10^4$ & $1.039 \times 10^4$ & $4.761 \times 10^6$ & 0.074 \\
$\cos \theta _{\mathrm{miss}}$ & 160.6 & 0.155 & 0.912 & 38.03 & 8719 & 1906 & $5.155 \times 10^5$ & 0.22 \\
$M_Z$ & 148.0 & 0 & 0.017 & 29.09 & 2408 & 501.2 & $1.299 \times 10^4$ & 1.2 \\
$\cos \theta _{e^- (e^+)}$ & 133.9 & 0 & 0.009 & 25.40 & 1067 & 101.5 & 729.7 & 3.0 \\
$E_{e^- (e^+)}$ & 133.0 & 0 & 0.009 & 24.93 & 690.3 & 78.70 & 629.7 & 3.4 \\
$\cos \theta _{\tau ^+ \tau ^-}$ & 130.8 & 0 & 0 & 3.536 & 254.9 & 30.70 & 155.4 & 5.5 \\
$\cos \theta _{\tau ^- (\tau ^+)}$ & 123.4 & 0 & 0 & 3.074 & 212.1 & 9.161 & 3.817 & 6.6 \\
$M_{\mathrm{recoil}}$ & 108.9 & 0 & 0 & 2.474 & 72.35 & 1.134 & 0.034 & 8.0 \\ \hline
\end{tabular}
\label{tab:Ztoee250_cuttable}
\end{table}

\begin{figure}[!h]
\centering
\includegraphics[width = 10.0truecm]{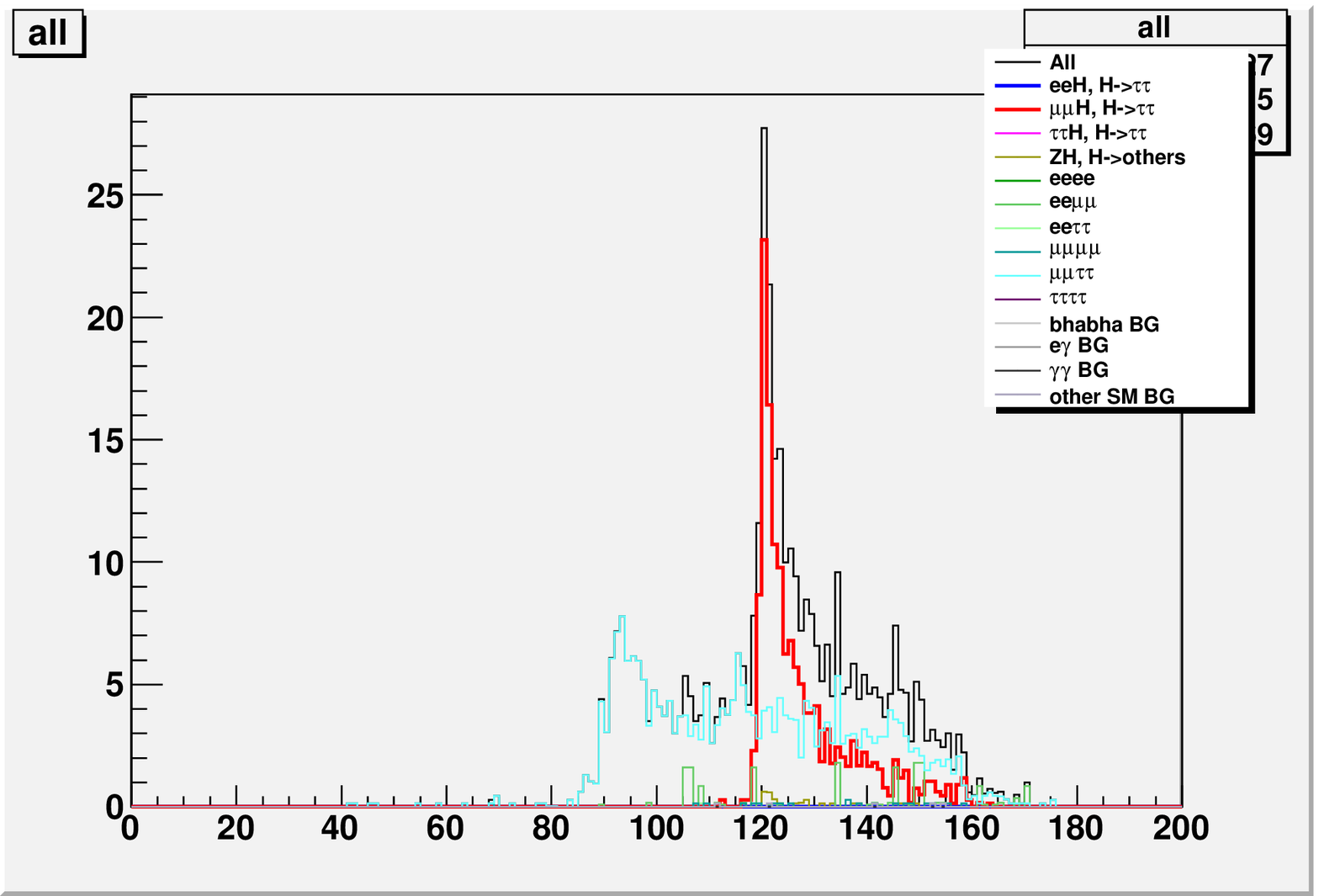}
\caption{The distribution of recoil mass in the unit of GeV.}
\label{Ztomumu250_recoilmass}
\end{figure}

We apply the following cuts sequentially for $Z \to \mu ^+ \mu ^-$ mode: number of tracks $\le 8$, 115 GeV $< E_{\mathrm{vis}} < 235$ GeV, $|\cos \theta _{\mathrm{miss}}| < 0.98$, 72 GeV $< M_Z < 107$ GeV, $E_{e^- (e^+)} < 90$ GeV, $\cos \theta _{\tau ^+ \tau ^-} < -0.5$, and 118 GeV $< M_{\mathrm{recoil}} < 143$ GeV.
Figure~\ref{Ztomumu250_recoilmass} shows the recoil mass distribution.
Table~\ref{tab:Ztomumu250_cuttable} shows the cut statistics of this mode.
For the $Z \to \mu ^+ \mu ^-$ mode case, 131.2 signal events and 91.2 background events are remained.
The significance is $S / \sqrt{S + B} = 131.2 / \sqrt{131.2 + 91.2} = 8.8 \sigma$.

\begin{table}[!h]
\centering
\caption{The cut statistics of $Z \to \mu ^+ \mu ^-$ mode at $\sqrt{s} = 250$ GeV.}
\begin{tabular}{c|cccccccc}\hline
\rule[-1pt]{0pt}{10pt} & $\mu ^+ \mu ^- h$ & $e^+ e^- h$ & $\tau ^+ \tau ^- h$ & $Zh$ with & $\mu ^+ \mu ^- \tau ^+ \tau ^-$ & other & other & signi. \\
 & $h \to \tau ^+ \tau ^-$ & $h \to \tau ^+ \tau ^-$ & $h \to \tau ^+ \tau ^-$ & $h \not \to \tau ^+ \tau ^-$ &  & 4 leptons & SM bkg. & \\ \hline
\rule[-1pt]{0pt}{10pt} No cut & 211.1 & 228.3 & 214.6 & 7325 & 3513 & $7.591 \times 10^6$ & $1.492 \times 10^{10}$ & 0.0017 \\
preselection & 168.5 & 0 & 0.155 & 43.01 & 1698 & 7546 & 7732 & 1.3 \\
\# of tracks & 167.4 & 0 & 0.155 & 39.65 & 1684 & 7537 & 7400 & 1.3 \\
$E_{\mathrm{vis}}$ & 162.9 & 0 & 0.155 & 37.40 & 1586 & 2285 & 3713 & 1.9 \\
$\cos \theta _{\mathrm{miss}}$ & 158.6 & 0 & 0.155 & 36.51 & 1386 & 227.5 & 55.48 & 3.7 \\
$M_Z$ & 153.2 & 0 & 0 & 32.84 & 1038 & 55.28 & 42.54 & 4.2 \\
$E_{e^- (e^+)}$ & 153.2 & 0 & 0 & 32.70 & 738.6 & 42.41 & 36.72 & 4.8 \\
$\cos \theta _{\tau ^+ \tau ^-}$ & 146.3 & 0 & 0 & 3.638 & 259.4 & 20.19 & 0.756 & 7.1 \\
$M_{\mathrm{recoil}}$ & 131.2 & 0 & 0 & 2.875 & 82.36 & 5.311 & 0.301 & 8.8 \\ \hline
\end{tabular}
\label{tab:Ztomumu250_cuttable}
\end{table}

\subsection{$Z \to q\overline{q}$ mode at $\sqrt{s} = 250$ GeV}

In this mode, the tau pairs are reconstructed first, followed by the dijet reconstruction of the $Z$ decay.

At first we apply the tau finder to all objects to reconstruct taus.
This tau finder searches the highest energy track and combine the neighboring particles, which satisfy $\cos \theta _{\mathrm{cone}} > 0.98$, with the combined mass less than 2 GeV.
We regard the combined object as a tau candidate.
Then we apply the selection cuts as following: $E_{\mathrm{tau \ candidate}} > 3$ GeV, $E_{\mathrm{cone}} < 0.1E_{\mathrm{tau \ candidate}}$ with $\cos \theta _{\mathrm{cone}} = 0.9$, and rejecting 3-prong with neutral particles events.
These selection cuts are tuned for minimizing misidentification of part of quark jets as tau jets.
A survived tau candidate is regarded as a tau jet.
After the selection cuts, we apply the charge recovery to obtain better efficiency.
The charged particles in tau jet which have the energy less than 2 GeV are detached one by one from smallest energy from the tau jet until satisfying the conditions as following: the charge of tau jet is $\pm 1$, and the number of track(s) in tau jet is 1 or 3.
The tau jet after detaching is rejected if it does not satisfy the above conditions.
After the selection cuts and detaching, we repeat the above processes until there are no charged particles which have the energy greater than 2 GeV.

\begin{figure}[!h]
\centering
\includegraphics[width = 10.0truecm]{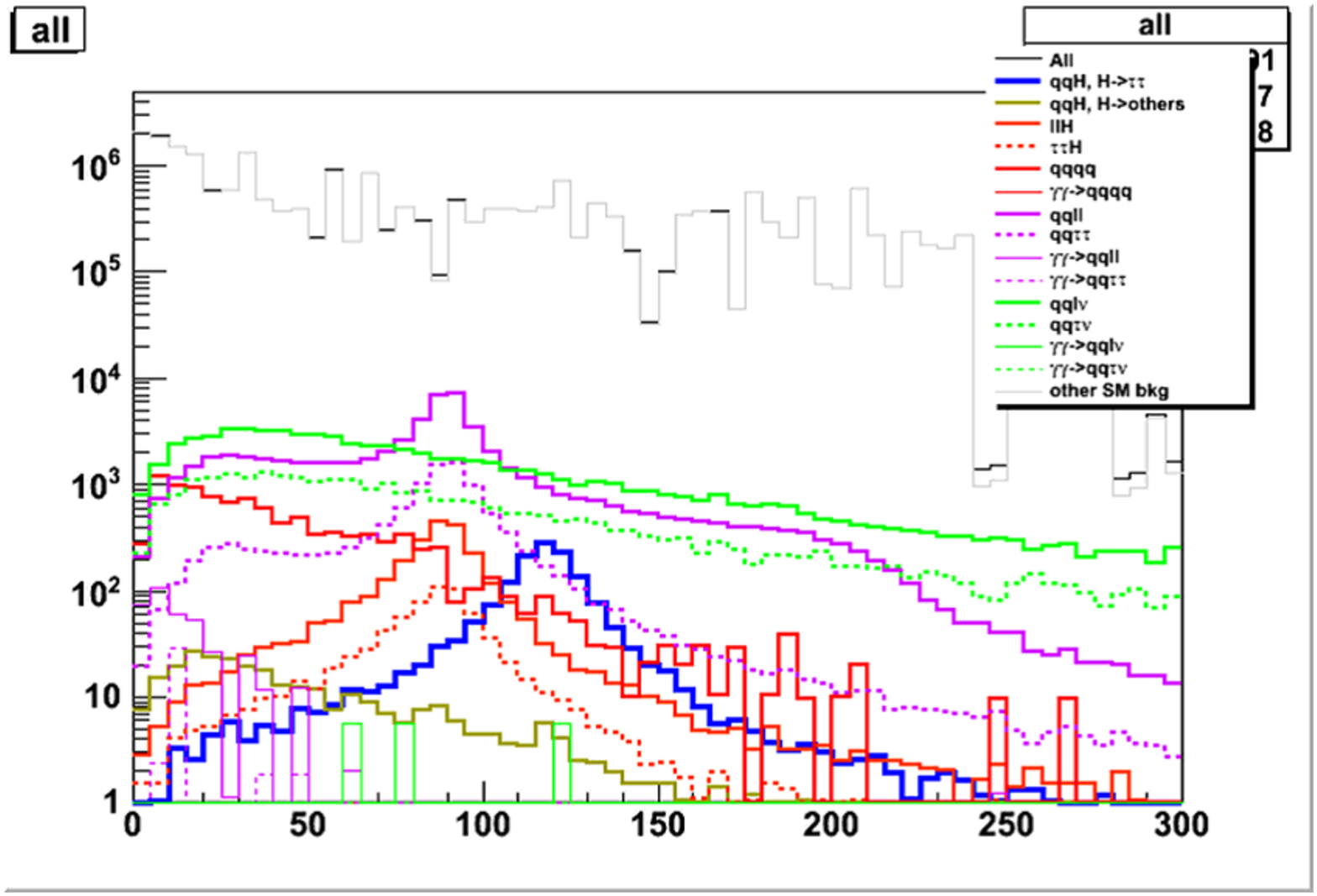}
\caption{The plot of $M_{\mathrm{colapp}}$ in the unit of GeV, the invariant mass of tau pair with collinear approximation.
Blue histogram shows the signal process $Zh \to q\overline{q}\tau ^+ \tau ^-$.}
\label{colapp}
\end{figure}

After the tau reconstruction, we apply the collinear approximation~\cite{colapp} to reconstruct tau pair.
In this approximation, we assume that the visible decay products of tau and the neutrino(s) from tau is collinear, and the contribution of missing transverse momentum is only comes from the neutrino(s) of tau decay.
The invariant mass of the tau pair with the collinear approximation shown in Figure~\ref{colapp}.

After that, we apply the Durham jet clustering method~\cite{Durham} with two jets for the remaining objects to reconstruct $Z$ boson.

\begin{figure}[!h]
\centering
\includegraphics[width = 10.0truecm]{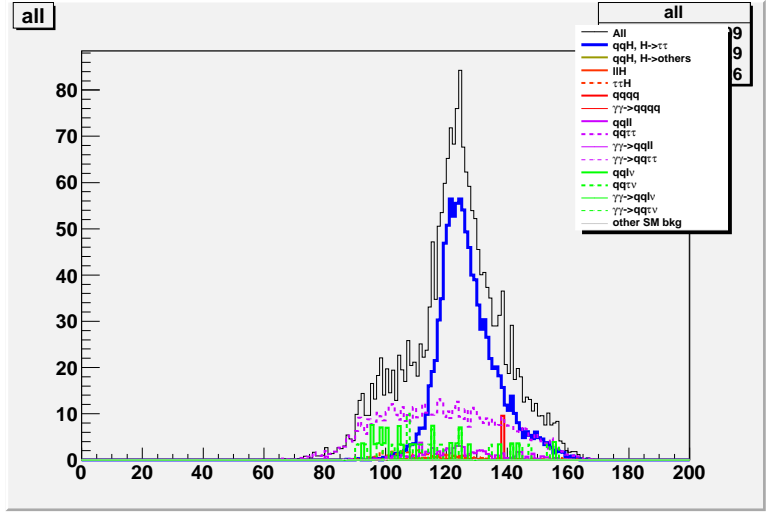}
\caption{The distribution of recoil mass in the unit of GeV.}
\label{Ztoqq250_recoilmass}
\end{figure}

After the all reconstruction, we apply the cuts to select signal process.
Before optimizing cuts, we apply the preselection as follows: number of quark jets $= 2$, number of $\tau ^+$ and $\tau ^- = 1$, number of tracks in a tau $\le 3$, and the events which have tracks in both taus $= 3$ are rejected (double 3-prong cut). 
We apply the following cuts sequentially to reject the background: $9 \le$ number of tracks $< 50$, 110 GeV $< E_{\mathrm{vis}} < 235$ GeV, $|\cos \theta _{\mathrm{miss}}| < 0.98$, 77 GeV $< M_Z < 135$ GeV, 80 GeV $< E_Z < 135$ GeV, $\cos \theta _{\tau ^+ \tau ^-} < -0.5$, $\log _{10}|d_0 / \sigma (d_0)|(\tau ^+) + \log _{10}|d_0 / \sigma (d_0)|(\tau ^-) > -0.7$, $\log _{10}|z_0 / \sigma (z_0)|(\tau ^+) + \log _{10}|z_0 / \sigma (z_0)|(\tau ^-) > -0.1$, $M_{\tau ^+ \tau ^-} < 115$ GeV, $E_{\tau ^+ \tau ^-} < 125$ GeV, 100 GeV $< M_{\mathrm{colapp}} < 170$ GeV, 100 GeV $< E_{\mathrm{colapp}} < 280$ GeV, and 112 GeV $< M_{\mathrm{recoil}} < 160$ GeV, where $M_{\tau ^+ \tau ^-}$ and $E_{\tau ^+ \tau ^-}$ is the invariant mass and energy of tau pair without using collinear approximation, $M_{\mathrm{colapp}}$ and $E_{\mathrm{colapp}}$ is the invariant mass and energy of tau pair with collinear approximation, respectively.
Figure~\ref{Ztoqq250_recoilmass} shows the distribution of recoil mass.
Table~\ref{tab:Ztoqq250_summary} shows the cut statistics of this mode.
After the cuts, the signal events and background events are remained 1026 and 554.4.
The statistical significance of $Z \to q\overline{q}$ mode is calculated to be $S / \sqrt{S + B} = 1026 / \sqrt{1026 + 554.4} = 25.8 \sigma$.

\begin{table}[!h]
\centering
\caption{The cut statistics of $Z \to q\overline{q}$ mode at $\sqrt{s} = 250$ GeV.}
{\tiny
\begin{tabular}{c|ccccccccccc}\hline
\rule[-1pt]{0pt}{8pt} & $q\overline{q}h$ & $Zh$ with & $\ell ^+ \ell ^- h$ & $\tau ^+ \tau ^- h$ & $q\overline{q}q\overline{q}$ & $q\overline{q}\ell ^+ \ell ^-$ & $q\overline{q}\tau ^+ \tau ^-$ & $q\overline{q}\ell \nu$ & $q\overline{q}\tau \nu$ & other & signi. \\
 & $h \to \tau ^+ \tau ^-$ & $h \not \to \tau ^+ \tau ^-$ &  &  &  &  &  &  & & SM bkg & \\ \hline
\rule[-1pt]{0pt}{8pt} No cut & 4233 & $4.829 \times 10^4$ & 5377 & 2596 & $4.038 \times 10^6$ & $3.563 \times 10^5$ & $4.169 \times 10^4$ & $2.788 \times 10^6$ & $1.326 \times 10^6$ & $1.494 \times 10^{10}$ & 0.035 \\
preselection & 1647 & 578.8 & 2761 & 765.4 & $1.230 \times 10^4$ & $6.378 \times 10^4$ & $1.161 \times 10^4$ & $1.249 \times 10^5$ & $4.948 \times 10^4$ & $2.570 \times 10^7$ & 0.32 \\
\# of tracks & 1644 & 549.8 & 2680 & 765.4 & $1.230 \times 10^4$ & $6.059 \times 10^4$ & $1.146 \times 10^4$ & $1.214 \times 10^5$ & $4.806 \times 10^4$ & $5.190 \times 10^5$ & 1.9 \\ 
$E_{\mathrm{vis}}$ & 1607 & 492.3 & 1015 & 744.2 & 4443 & $2.106 \times 10^4$ & $1.107 \times 10^4$ & $1.192 \times 10^5$ & $4.693 \times 10^4$ & $2.383 \times 10^5$ & 2.4 \\
$\cos \theta _{\mathrm{miss}}$ & 1572 & 474.7 & 860.5 & 725.1 & 2127 & 8315 & $1.021 \times 10^4$ & $1.171 \times 10^5$ & $4.415 \times 10^4$ & 5939 & 3.6 \\
$M_Z$ & 1440 & 376.1 & 791.3 & 682.8 & 778.6 & 4987 & 8674 & 8189 & 3288 & 997.3 & 8.3 \\
$E_Z$ & 1429 & 352.0 & 782.7 & 528.7 & 505.0 & 4797 & 7857 & 7703 & 3061 & 609.9 & 8.6 \\
$\cos \theta _{\tau ^+ \tau ^-}$ & 1386 & 46.28 & 442.2 & 255.6 & 191.4 & 1468 & 2001 & 2831 & 1154 & 475.6 & 13.7 \\
$d_0\mathrm{sig}$ & 1338 & 30.29 & 235.1 & 244.3 & 131.4 & 854.9 & 1928 & 1786 & 1044 & 248.1 & 15.1 \\
$z_0\mathrm{sig}$ & 1287 & 19.54 & 105.0 & 234.7 & 81.77 & 408.2 & 1845 & 909.9 & 883.4 & 244.6 & 16.6 \\
$M_{\tau ^+ \tau ^-}$ & 1286 & 19.39 & 103.2 & 234.7 & 72.05 & 349.1 & 1837 & 883.5 & 883.4 & 243.9 & 16.7 \\
$E_{\tau ^+ \tau ^-}$ & 1282 & 19.39 & 103.0 & 234.7 & 72.05 & 324.7 & 1836 & 873.2 & 883.4 & 243.9 & 16.7 \\
$M_{\mathrm{colapp}}$ & 1065 & 3.074 & 18.76 & 47.94 & 10.28 & 72.83 & 616.9 & 150.8 & 137.0 & 0.746 & 23.1 \\
$E_{\mathrm{colapp}}$ & 1062 & 2.454 & 18.01 & 46.72 & 10.28 & 71.27 & 612.1 & 93.05 & 93.52 & 0.454 & 23.7 \\
$M_{\mathrm{recoil}}$ & 1026 & 2.144 & 14.54 & 21.24 & 9.938 & 57.07 & 366.3 & 39.64 & 43.31 & 0.161 & 25.8 \\ \hline
\end{tabular}
}
\label{tab:Ztoqq250_summary}
\end{table}

\subsection{$Z \to q\overline{q}$ mode at $\sqrt{s} = 500$ GeV}

\begin{figure}[!h]
\centering
\includegraphics[width = 10.0truecm]{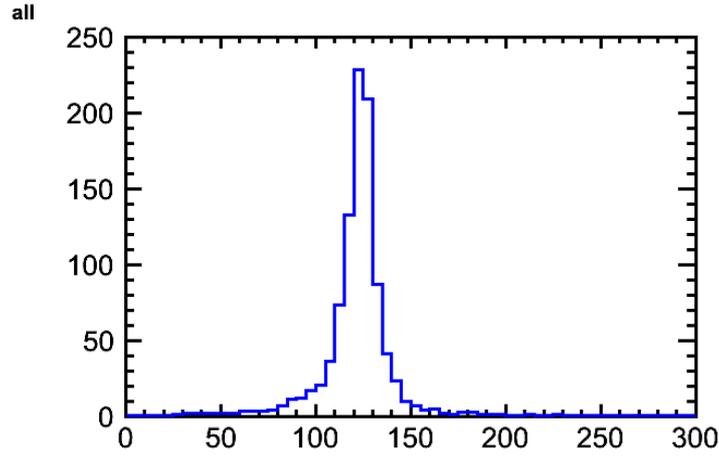}
\caption{The plot of $M_{\mathrm{colapp}}$ of the signal process $Zh \to q\overline{q}\tau ^+ \tau ^-$ in the unit of GeV, the invariant mass of tau pair with collinear approximation.}
\label{colapp_500_qq}
\end{figure}

We take the same analysis strategy which described in Section 4.2.
We apply the same tau finder to all objects to reconstruct taus from Higgs boson, followed by collinear approximation~\cite{colapp}, then apply Durham algorithm~\cite{Durham} to remaining objects to reconstruct $Z$ boson.
Figure~\ref{colapp_500_qq} shows the distribution of tau pair mass with collinear approximation for the signal process.

\begin{figure}[!h]
\centering
\includegraphics[width = 10.0truecm]{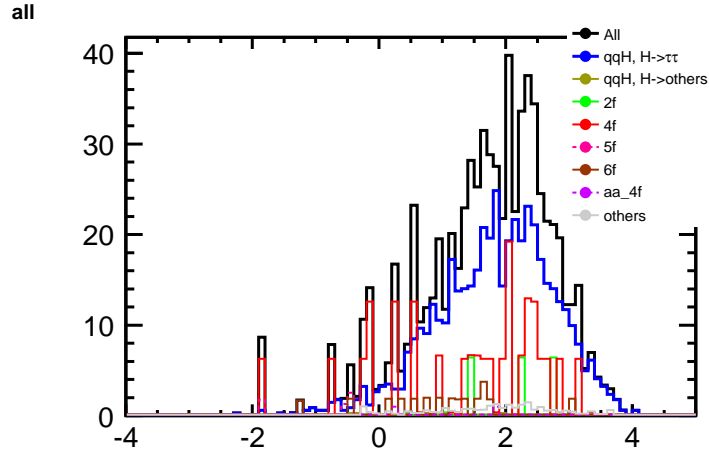}
\caption{The distribution of $\log _{10}|z_0 / \sigma (z_0)|(\tau ^+) + \log _{10}|z_0 / \sigma (z_0)|(\tau ^-)$.}
\label{qqh500_z0sig}
\end{figure}

After the reconstruction we apply the cuts to select signal process.
Before optimizing, we apply the preselection cut as; number of quark jet = 2, number of $\tau ^+ (\tau ^-) = 1$, number of tracks in a tau $\le 3$.
Then we apply following cuts: number of tracks $\ge 14$, thrust $< 0.93$, $|\cos \theta _{\mathrm{miss}}| < 0.95$, 70 GeV $< M_Z < 265$ GeV, $E_Z > 135$ GeV, 20 GeV $< M_{\tau ^+ \tau ^-} < 120$ GeV, $E_{\tau ^+ \tau ^-} < 235$ GeV, $\cos \theta _{\tau ^+ \tau ^-} < 0.56$, 115 GeV $< M_{\mathrm{colapp}} < 135$ GeV, 205 GeV $< E_{\mathrm{colapp}} < 270$ GeV, $\log _{10}|d_0 / \sigma (d_0)|(\tau ^+) + \log _{10}|d_0 / \sigma (d_0)|(\tau ^-) > 0.4$, and $\log _{10}|z_0 / \sigma (z_0)|(\tau ^+) + \log _{10}|z_0 / \sigma (z_0)|(\tau ^-) > -0.1$.
Figure~\ref{qqh500_z0sig} shows the distribution of $\log _{10}|z_0 / \sigma (z_0)|(\tau ^+) + \log _{10}|z_0 / \sigma (z_0)|(\tau ^-)$.
Table~\ref{tab:Ztoqq500_summary} shows the cut statistics of this mode.
The statistical significance of this mode is calculated to be $S / \sqrt{S + B} = 453.7 / \sqrt{453.7 + 219.5} = 17.5 \sigma$.
This result corresponds to the precision of $\Delta (\sigma \cdot \mathrm{Br}) / (\sigma \cdot \mathrm{Br}) = 5.7 \ \%$

\begin{table}[!h]
\centering
\caption{The cut statistics of $Z \to q\overline{q}$ mode at $\sqrt{s} = 500$ GeV.}
{\scriptsize
\begin{tabular}{c|ccccccccc}\hline
\rule[-1pt]{0pt}{8pt} & $q\overline{q}h$ & $q\overline{q}h$ with & 2f & 4f & 5f & 6f & $\gamma \gamma \to$ 4f & $\ell ^+ \ell ^- h$, $\nu \overline{\nu} h$ & signi. \\
 & $h \to \tau ^+ \tau ^-$ & $h \not \to \tau ^+ \tau ^-$ &  &  &  &  & & & \\ \hline
\rule[-1pt]{0pt}{8pt} No cut & 2158 & $3.139 \times 10^4$ & $1.320 \times 10^7$ & $1.598 \times 10^7$ & $6.895 \times 10^4$ & $5.887 \times 10^5$ & $1.041 \times 10^5$ & $9.510 \times 10^4$ & 0.39 \\
preselection & 1019 & 604.9 & $1.151 \times 10^6$ & $1.125 \times 10^6$ & 9713 & $4.001 \times 10^4$ & $1.349 \times 10^4$ & 7694 & 0.67 \\
\# of tracks & 994.5 & 600.8 & $1.569 \times 10^5$ & $3.498 \times 10^5$ & 6960 & $3.879 \times 10^4$ & 5187 & 5403 & 1.3 \\ 
thrust & 964.1 & 574.5 & $7.176 \times 10^4$ & $1.821 \times 10^5$ & 6663 & $3.852 \times 10^4$ & 4822 & 5821 & 1.7 \\
$\cos \theta _{\mathrm{miss}}$ & 898.3 & 486.4 & $1.741 \times 10^4$ & $1.174 \times 10^5$ & 4195 & $3.508 \times 10^4$ & 2379 & 3955 & 2.1 \\
$M_Z$ & 855.3 & 321.2 & 7682 & $7.962 \times 10^4$ & 3467 & $2.488 \times 10^4$ & 1741 & 3441 & 2.5 \\
$E_Z$ & 849.7 & 318.5 & 5164 & $6.666 \times 10^4$ & 2884 & $2.282 \times 10^4$ & 1243 & 3147 & 2.7 \\
$M_{\tau ^+ \tau ^-}$ & 806.1 & 259.4 & 1530 & $4.243 \times 10^4$ & 1483 & $1.687 \times 10^4$ & 665.4 & 1974 & 3.1 \\
$E_{\tau ^+ \tau ^-}$ & 800.3 & 258.0 & 1154 & $3.465 \times 10^4$ & 1436 & $1.684 \times 10^4$ & 645.7 & 1112 & 3.4 \\
$\cos \theta _{\tau ^+ \tau ^-}$ & 795.9 & 137.5 & 744.6 & $2.699 \times 10^4$ & 1093 & $1.471 \times 10^4$ & 490.4 & 472.8 & 3.7 \\
$M_{\mathrm{colapp}}$ & 557.8 & 6.435 & 52.91 & 770.4 & 38.19 & 579.6 & 17.78 & 36.81 &  12.3 \\
$E_{\mathrm{colapp}}$ & 511.3 & 5.265 & 38.91 & 351.5 & 20.44 & 90.13 & 7.943 & 31.67 &  15.7 \\
$d_0\mathrm{sig}$ & 468.6 & 2.047 & 20.44 & 179.9 & 5.623 & 28.35 & 1.995 & 24.30 & 17.3 \\
$z_0\mathrm{sig}$ & 453.7 & 1.462 & 20.44 & 148.4 & 0 & 24.57 & 1.995 & 22.58 & 17.5 \\ \hline
\end{tabular}
}
\label{tab:Ztoqq500_summary}
\end{table}

\subsection{$WW$-fusion process at $\sqrt{s} = 500$ GeV}

At first in this mode, we apply the $k_T$ algorithm~\cite{kT1, kT2} to remove objects from $\gamma \gamma \to$ hadron(s) overlaid process.
We use \verb|FastJet| package~\cite{FastJet} as the $k_T$ clustering package.
We choose the value of generalized radius $R$ of $k_T$ clustering of 1.0 currently (more optimization needed).

We apply the same tau finder which described in Section 4.1 to all survived objects through the $k_T$ clustering, the only difference is the maximum associated angle has been changed from 1.0 radian to 0.76 radian.
The most energetic $\tau ^+$ candidate and $\tau ^-$ candidate are combined as it comes from Higgs boson.

After the reconstruction, we apply the preselection as the number of $\tau ^+ (\tau ^-) \ge 1$, because the $\gamma \gamma \to$ hadron(s) processes produce additional charged particles, and the tau finder which used for this process repeat finding process until there are no charged particles.

\begin{figure}[!h]
\centering
\includegraphics[width = 10.0truecm]{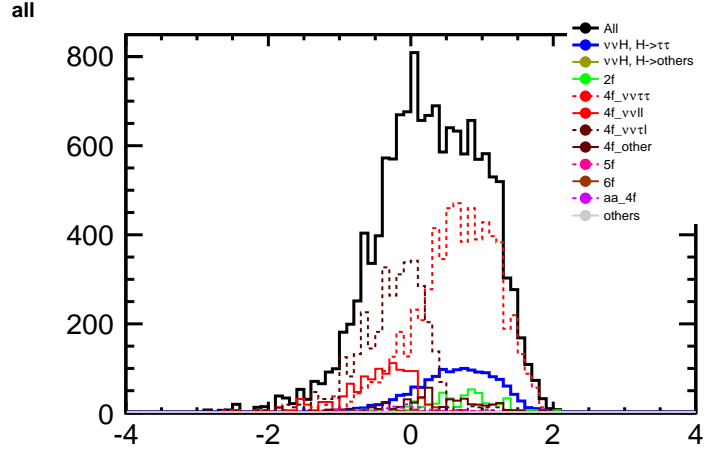}
\caption{The distribution of $(\min(z_0 / \sigma (z_0)))$.}
\label{nunuh500_z0sig}
\end{figure}

Then we apply following cuts: number of tracks $\le 6$, 5 GeV $< M_{\mathrm{vis}} < 135$ GeV, $E_{\mathrm{vis}} < 240$ GeV, $P_t > 25$ GeV, $|\cos \theta _{\mathrm{miss}}| < 0.89$, $M_{\tau ^+ \tau ^-} < 115$ GeV, $-0.86 < \cos \theta _{\tau ^+ \tau ^-} < 0.57$, $\cos \theta _{\mathrm{acop}} < 0.99$, $\log _{10}|\min (d_0 / \sigma (d_0))| > 0.3$, $\log _{10}|\min (z_0 / \sigma (z_0))| > 0$, where $M_{\mathrm{vis}}$ is visible energy, $\theta _{\mathrm{acop}}$ is acoplanarity, $\min(d_0 / \sigma (d_0))$ $(\min(z_0 / \sigma (z_0)))$ is smaller impact parameter value between $\tau ^+$ and $\tau ^-$, respectively.
Figure~\ref{nunuh500_z0sig} shows the distribution of $\log _{10} (\min(z_0 / \sigma (z_0)))$.
Table~\ref{tab:nunuh500_summary} shows the cut statistics of this mode.
The statistical significance of this mode is calculated to be $S / \sqrt{S + B} = 1469 / \sqrt{1469 + 1.061 \times 10^4} = 13.4 \sigma$.

\begin{table}[!h]
\centering
\caption{The cut statistics of $WW$-fusion process at $\sqrt{s} = 500$ GeV.}
{\tiny
\begin{tabular}{c|cccccccccc}\hline
\rule[-1pt]{0pt}{8pt} & $\nu \overline{\nu} h$ & $\nu \overline{\nu} h$ with & 2f & 4f & 4f & 5f & 6f & $\gamma \gamma \to$ 4f & $\ell ^+ \ell ^- h$, $q\overline{q}h$ & signi. \\
 & $h \to \tau ^+ \tau ^-$ & $h \not \to \tau ^+ \tau ^-$ &  & $\nu \overline{\nu} \tau ^+ \tau ^-$ & others &  & & & \\ \hline
\rule[-1pt]{0pt}{8pt} No cut & 5401 & $7.967 \times 10^4$ & $1.335 \times 10^7$ & $1.452 \times 10^7$ & $1.594 \times 10^7$ & $6.895 \times 10^4$ & $5.895 \times 10^5$ & $1.041 \times 10^5$ & $4.363 \times 10^4$ & 0.98 \\
preselection & 4676 & 6062 & $3.973 \times 10^6$ & $1.070 \times 10^5$ & $4.215 \times 10^6$ & $2.093 \times 10^4$ & $2.344 \times 10^4$ & $4.132 \times 10^4$ & 2650 & 1.6 \\
\# of tracks & 4274 & 2461 & $2.649 \times 10^6$ & $9.896 \times 10^4$ & $2.457 \times 10^6$ & $1.123 \times 10^4$ & 7035 & $2.782 \times 10^4$ & 1126 & 1.9 \\
$M_\mathrm{vis}$ & 4258 & 2394 & $8.030 \times 10^5$ & $5.527 \times 10^4$ & $1.087 \times 10^6$ & 7220 & 2773 & $2.458 \times 10^4$ & 262.9 & 3.0 \\ 
$E_{\mathrm{vis}}$ & 4251 & 2386 & $5.846 \times 10^5$ & $5.511 \times 10^4$ & $9.096 \times 10^5$ & 6886 & 2736 & $2.357 \times 10^4$ & 205.9 & 3.4 \\
$P_t$ & 3992 & 2166 & $4.232 \times 10^5$ & $4.580 \times 10^4$ & $5.043 \times 10^5$ & 5229 & 2683 & 9244 & 205.9 & 4.0 \\
$\cos \theta _{\mathrm{miss}}$ & 3294 & 1876 & $6.745 \times 10^4$ & $3.051 \times 10^4$ & $1.967 \times 10^5$ & 2702 & 1978 & 4368 & 148.7 & 5.9 \\
$M_{\tau ^+ \tau ^-}$ & 3245 & 1865 & $5.612 \times 10^4$ & $2.653 \times 10^4$ & $1.759 \times 10^5$ & 2485 & 1696 & 4157 & 141.7 & 6.2 \\
$\cos \theta _{\tau ^+ \tau ^-}$ & 2837 & 923.6 & $1.495 \times 10^4$ & $1.416 \times 10^4$ & $1.085 \times 10^5$ & 1757 & 1224 & 2866 & 64.59 & 7.4 \\
$\cos \theta _{\mathrm{acop}}$ & 2742 & 909.3 & 7384 & $1.367 \times 10^4$ & $1.050 \times 10^5$ & 1722 & 1201 & 2792 & 63.65 & 7.5 \\
$d_0\mathrm{sig}$ & 1733 & 77.50 & 745.2 & 8293 & 8051 & 159.9 & 134.0 & 261.7 & 11.71 & 12.4 \\
$z_0\mathrm{sig}$ & 1469 & 40.49 & 542.5 & 6989 & 2744 & 84.90 & 76.22 & 126.4 & 7.755 & 13.4 \\
\hline
\end{tabular}
}
\label{tab:nunuh500_summary}
\end{table}

In $\nu \overline{\nu} h$ events, there are two contributions from $WW$-fusion and Higgs-strahlung.
The number of remained events can be written as:
\begin{equation*}
N_{\mathrm{remained}} = L \left( \sum _{i = e, \mu, \tau} \sigma _{Zh} \times \mathrm{Br} (Z \to \nu _i \overline{\nu _i}) \times \mathrm{Br} (h \to \tau ^+ \tau ^-) \times \varepsilon _1 + \sigma _{WW\mathrm{-fusion}} \times \mathrm{Br} (h \to \tau ^+ \tau ^-) \times \varepsilon _2 \right) ,
\end{equation*}
where $\varepsilon _1$ and $\varepsilon _2$ are the selection efficiency for Higgs-strahlung process and $WW$-fusion process, respectively.
The signal significance is for the $\sum _{i = e, \mu \tau} \sigma _{Zh} \times \mathrm{Br} (Z \to \nu _i \overline{\nu _i}) \times \mathrm{Br} (h \to \tau ^+ \tau ^-) + \sigma _{WW\mathrm{-fusion}} \times \mathrm{Br} (h \to \tau ^+ \tau ^-)$.

\section{Summary and Prospects}

We evaluate the measurement accuracy of the branching ratio of the $h \to \tau ^+ \tau ^-$ mode at $\sqrt{s} = 250$ GeV and 500 GeV at the ILC with ILD detector full simulation.
For the analysis of $\sqrt{s} = 250$ GeV, we assume $M_h = 120$ GeV, $\mathrm{Br}(h \to \tau ^+ \tau ^-) = 8.0 \ \%$, $\int L \ dt = 250 \ \mathrm{fb^{-1}}$, and beam polarization $P(e^+, e^-) = (+0.3, -0.8)$.
The analysis results and scaled results to $M_h = 125$ GeV are summarized in Table \ref{250GeV_summary}.

\begin{table}[!h]
\centering
\caption{The analysis results of $\sqrt{s} = 250$ GeV with assuming $M_h = 120$ GeV and scaled results to $M_h = 125$ GeV.}
\begin{tabular}{cccc|c|c}
\hline
 & $Z \to e^+ e^-$ & $Z \to \mu ^+ \mu ^-$ & $Z \to q\overline{q}$ & Combined & $\dfrac{\Delta (\sigma \cdot \mathrm{Br})}{(\sigma \cdot \mathrm{Br})}$ \\
\hline
Results of $M_h = 120$ GeV & $8.0 \sigma$ & $8.8 \sigma$ & $25.8 \sigma$ & $28.4 \sigma$ & $3.5 \ \%$ \\
Scaled results to $M_h = 125$ GeV & $6.8 \sigma$ & $7.4 \sigma$ & $21.9 \sigma$ & $24.1 \sigma$ & $4.2 \ \%$ \\
\hline
\label{250GeV_summary}
\end{tabular}
\end{table}

For the $\sqrt{s} = 500$ GeV, the analyses are still ongoing, but we obtain the statistical significance and measurement accuracy as summarized in Table~\ref{500GeV_summary}, with assuming $M_h = 125$ GeV, $\mathrm{Br}(h \to \tau ^+ \tau ^-) = 6.32 \ \%$, $\int L \ dt = 500 \ \mathrm{fb^{-1}}$, and beam polarization $P(e^+, e^-) = (+0.3, -0.8)$.
The result of $WW$-fusion is better than the expected accuracy in the Technical Design Report~\cite{TDR2}.
We expect improvement by better treatment of $\gamma \gamma \to$ hadron(s) background and more optimizing tau reconstruction.

\begin{table}[!h]
\centering
\caption{The analysis results of $\sqrt{s} = 500$ GeV with assuming $M_h = 125$ GeV.}
\begin{tabular}{ccc}
\hline
 & $Z \to q\overline{q}$ & $WW$-fusion \\
\hline
significance & $17.5 \sigma$ & $13.4 \sigma$ \\
\hline
$\dfrac{\Delta (\sigma \cdot \mathrm{Br})}{(\sigma \cdot \mathrm{Br})}$ & $5.7 \ \%$ & $7.5 \ \%$ \\
\hline
\label{500GeV_summary}
\end{tabular}
\end{table}

\end{document}